\documentclass[11pt]{article}   	
\usepackage{geometry}                		
\usepackage{graphicx}				
\usepackage{amssymb}
\usepackage{amsmath}
\usepackage{scalerel}
\usepackage{cite}
\usepackage{authblk}

\usepackage{color}

\definecolor{ascolor}{rgb}{1,0,1}

\newcommand\asout{\marginpar{\color{ascolor}$\heartsuit$}\bgroup\markoverwith{\color{ascolor}{\rule[0.4ex]{2pt}{0.8pt}}}\ULon}

\def\asbar{\bar{\alpha}_s}

\newcommand{\ci}{{\mathrm{c}}}
\newcommand{\be}{\begin{eqnarray}}
\newcommand{\ee}{\end{eqnarray}}

\newcommand{\bea}{\begin{align}}
\newcommand{\eea}{\end{align}}

\newcommand{\CK}{{ K}}

\newcommand{\kt}{{\mathbf{k}}}       
\newcommand{\qt}{\mathbf{q}}        
\newcommand{\Li}{{\mathrm{Li}_2}}       

\newcommand{\kc}{\mathrm{kc}}           

\title{\bf Taming  of preasymptotic small $x$ evolution within resummation framework  }
\author[1]{Michal Deak}
\author[1,2]{Leonid Frankfurt}
\author[1]{Mark Strikman}
\author[1]{ Anna M. Sta\'sto}

\affil[1]{\small \it Department of Physics, Penn State University, University Park, PA 16802, USA}
\affil[2]{\small \it Sackler School of Exact Sciences, Tel Aviv University, Tel Aviv, 69978, Israel}
\begin{document}
\maketitle

\begin{abstract}
It is well understood  that the leading logarithmic approximation for the amplitudes of high energy processes is insufficient and that the next-to-leading logarithmic effects are very large and lead to instability of the solution. The  resummation at low $x$, which includes kinematical constraints and other corrections  leads to stable result.
Using previously established resummation procedure we study in detail the preasymptotic effects which occur in the solution to the resummed BFKL equation when the energy is not very large. We find that in addition to the well known reduction of the intercept, which governs the energy dependence of the gluon Green's function, resummation leads to the delay of the onset of its small $x$ growth. Moreover the gluon Green's function develops a dip or a plateau in wide range of rapidities, which increases for large scales. The preasymptotic region  in the gluon Green's function extends to about  $8$ units  in rapidity for  the  transverse scales  of the  order  of $30-100 \; {\rm GeV} $. 
 To visualize the expected behavior of physical  processes  with two equal hard scales we  calculate the cross section of the process $\gamma^{*}+\gamma^{*}\to X$  to be probed at future very high-energy electron-positron colliders.  We find that at $\gamma^*\gamma^*$ energies below $100 \; \rm GeV$ the BFKL Pomeron leads to  smaller value of the cross section than the Born approximation, and only starts to dominate  at  energies about $100 \; \rm GeV$.
 This pattern is significantly different from the one which we find using LL approximation.
  We also analyze the transverse momentum contributions to the cross section for different virtualities of the photons and find that the dominant contributions to the integral over the transverse momenta comes from lower values than the the external scales in the process under consideration. 
\end{abstract}

\maketitle


\section{Introduction and Motivation}
\label{sec:intro}

One of the outstanding and interesting problems in the domain of strong interactions is the behavior of the cross sections of hard processes in QCD in the high energy limit.
High energy Deep Inelastic Scattering experiment at HERA established very strong growth of the structure functions  with decreasing values of Bjorken $x$ \cite{Derrick:1993fta,Abt:1993cb}.
This growth was subsequently understood as a result of the increase of the gluon density at low $x$ due to the non-abelian gluon splitting.
The increase of the parton densities towards low $x$ is also responsible for the growth of the various cross sections of hard processes at the LHC. 
The standard approach to the description of the evolution of parton densities is provided by the collinear picture and the  Dokshitzer-Gribov-Lipatov-Altarelli-Parisi (DGLAP) evolution equations \cite{Dokshitzer:1977sg,Gribov:1972ri,Altarelli:1977zs}. These equations have been derived under the assumption of fixed $x$ and very large scale $Q^2$, such that they sum
all  the powers of
$(\alpha_s)^{m} (\log (Q^2/\Lambda^2))^n$,
where $\Lambda$ is some hadronic scale.  The validity of the DGLAP equations 
is well established and the collinear parton densities are at the core of most of the perturbative predictions at the Large Hadron Collider.

On the other hand, the Balitsky-Fadin-Kuraev-Lipatov (BFKL) \cite{Kuraev:1977fs,Balitsky:1978ic, Lipatov:1985uk} equation
accounts for the evolution in $x$ at small values of $x$.  
Within this approximation, one assumes 
large center of mass energy $s$, which is much larger than any of the scales in the process, including the momentum transfer $t$ 
and hard scale $Q^2$.  
BFKL hard-Pomeron approximation is suitably defined for 
the processes which are characterized by the presence of two comparable, perturbative, scales, with a large rapidity distance in between. 
For such processes the evolution to infrared pole, to the non-perturbative regime where pQCD becomes doubtful, is kinematically suppressed.  To give few examples, such configurations can be realized in the production of Muller-Navelet jets in hadron-hadron scattering \cite{Mueller:1986ey}, forward jet production in electron-hadron DIS (where jet transverse momentum is comparable to photon virtuality) \cite{Mueller:1989mx,Mueller:1990gy} and in the $\gamma^*\gamma^*$ scattering in $e^+e^-$ collisions \cite{Brodsky:1996sg,Brodsky:1997sd}. In these processes the DGLAP configurations, which are strongly ordered in transverse momenta, are suppressed since virtualities in the projectile and the target fragmentation regions are comparable. But the transverse momenta of gluons  exchanged  in the $t$ channel are not ordered.  
It is thus expected that the BFKL evolution would  be the dominant mechanism for the growth of the cross section for such process with the energy. The 
LL (leading-logarithmic) BFKL equation resums terms  $(\alpha_s \ln s)^n$ which in turn leads to the power like 
increase of the cross sections with the center-of-mass energy $\sqrt{s}$.

The BFKL Pomeron in the leading logarithmic  approximation by definition violates energy-momentum conservation.   
A related effect was observed in QED in the process of production of $e^+e^-$ pairs by Lipatov and Kuraev  \cite{Kuraev:1973sj}.
For further discussion, see \cite{Frankfurt:2013ria}.
 This results in NLL corrections being very large and negative. Thus LL approximation is unstable and the urgent question arises on the necessity to guarantee stabilization of BFKL Pomeron approximation. BFKL Pomeron is currently available 
up to next-to-leading logarithmic accuracy in QCD \cite{Fadin:1998py,Ciafaloni:1998gs} and up to next-to-next-to-logarithmic accuracy in $N=4$ supersymmetric Yang-Mills theory (sYM) \cite{caron2018high, Gromov:2015vua, Velizhanin:2015xsa}. The corrections at the NLL level turned out to be very large, reducing the value of the intercept significantly and also leading to the instabilities like the oscillating cross sections 
including negative values.

Therefore it was soon realized that the resummation is necessary to stabilize the result.
The resummation program was developed, starting from the important observation that kinematical constraints yield large corrections when imposed onto the BFKL kernel \cite{Andersson:1995ju,Kwiecinski:1996td,Kwiecinski:1997ee}. Full resummation which included matching to NLL BFKL and appropriate subtractions was developed in series of works    \cite{Salam:1998tj, Salam:1999cn, Ciafaloni:1999au,Ciafaloni:1999yw,Ciafaloni:2003ek,Ciafaloni:2003rd,Ciafaloni:2003kd,Ciafaloni:2007gf},  \cite{Altarelli:1999vw,Altarelli:2000mh,Altarelli:2001ji,Altarelli:2003hk,Altarelli:2008aj}, \cite{Thorne:2001nr,White:2006yh} and  \cite{Vera:2005jt}.  The next-to-leading order corrections were also calculated \cite{Balitsky:2008zza,Kovner:2014lca,Lublinsky:2016meo} in the context of the non-linear evolution equation for high parton densities \cite{Kovchegov:1999ua,Kovchegov:1999yj,Balitsky:1995ub,Iancu:2000hn,Ferreiro:2001qy,JalilianMarian:1997dw,JalilianMarian:1997gr,JalilianMarian:1997jx} and turned out to be also very large. The resummation of higher order corrections was also applied  in this  context of the non-linear evolution  \cite{Motyka:2009gi,Beuf:2014uia,Iancu:2015joa,Iancu:2015vea}.  In  \cite{Bonvini:2016wki} the resummation formalism was applied to DIS process for the description of the  HERA data and was shown to improve the quality of fits with respect to the ones based on the pure DGLAP evolution.

In this paper we shall focus on the Ciafaloni-Colferai-Salam-Stasto (CCSS) resummation \cite{Ciafaloni:2003ek,Ciafaloni:2003rd,Ciafaloni:2003kd,Ciafaloni:2007gf}
 and its application to phenomenology. 
The main sources of the large NLL corrections to BFKL were identified as originating from the so-called double transverse logarithms, or in Mellin space representation of the kernel, from the triple collinear poles \cite{Salam:1998tj}. Large negative terms are also coming  from the parts of the leading order DGLAP splitting function which appear at NLL BFKL, and they manifest themselves as the double collinear poles. The third source of the large corrections at NLL BFKL is the running of the strong coupling. The strong coupling is fixed in LL BFKL and practically  only starts to run at NLL level 
\cite{Fadin:1998py}. This leads to the large correction, and the increased importance of the infrared regime. Even if the BFKL evolution is considered for the one scale process, like Mueller-Navelet jets, with two equal, large, perturbative scales, the diffusion of the transverse momenta along the ladder  leads to the increased importance of the infrared regime, where the coupling becomes very large.\footnote{In this paper we refer to processes with two equal scales as one-scale processes, and to processes with two unequal scales as two-scale processes.}
The CCSS resummation procedure was based on combining the LL and NLL BFKL together with kinematical constraint and DGLAP anomalous dimension. Appropriate subtractions needed to be performed in order to avoid the double counting and guarantee that the momentum sum rule was satisfied. The resummation procedure was shown to stabilize the result for the BFKL gluon Green's function and significantly reduced the value of the intercept from the LL result for wide range of values of $\alpha_s$.

 Additional important feature of resummation, which is not often stressed,  is the delay in the onset of power like increase, typical of BFKL resummation to lower values of $x$ or consequently  higher values of rapidity. This was first discussed in the context of the `dip' of the splitting function with resummation \cite{Ciafaloni:2003kd}.  It was demonstrated that the resummed splitting function exhibits a decrease in $x$ followed by the increase towards asymptotia. It was understood as the interplay between different terms in the perturbative expansion, notably between the terms which appear at NLL BFKL (or NNLO DGLAP splitting function) and the terms which appear at LL BFKL (and would appear at N$^3$LO DGLAP). In the region of moderate
values of $x$ the splitting function had a small dip, and thus could well be approximated by the flat function. 

This may partly explain the success of DGLAP in the description of the current HERA data on structure functions \cite{Bonvini:2016wki} and the fact that NLO fits seem to perform better than NNLO fits, since the $xP_{gg}$ splitting function at NLO is flat as a function of $x$. The region in which the BFKL growth starts to dominate is pushed towards smaller values of $x$, leading to relatively large flat region in $xP_{gg}$ function. The question thus remains, as to what is the domain in which the BFKL resummation becomes dominant and DGLAP dynamics ceases to become sufficient as a description of the high energy scattering.

The goal of this paper is to  investigate
in more detail the preasymptotic regime in the resummed BFKL aproach in the context of the one-scale process, like $\gamma^*\gamma^*$ scattering. In particular we want to establish quantitatively the boundary in  plane of rapidity and a hard scale at which the BFKL increase starts to occur and leads to the sharp of the cross section which is the signature of the BFKL regime.  We shall base our analysis on the CCSS resummation framework. First, we shall recall the solutions to the gluon Green's functions $G(Y;k_{0},k_{0})$ using CCSS resummation and perform the study on the level of the BFKL solution itself. We find that the length of the plateau in rapidity depends almost linearly on the logarithm of the scales. The BFKL growth dominates only for rather large rapidity intervals, starting from about $4$ units of rapidity for $k_0=10\,  \rm GeV$  to about $8$ units of rapidity for $k_0=100\, \rm GeV$.  Second, for the more realistic case, we shall then apply this solution to the process with comparable hard external scales. Such processes were argued to be the gold-plated signatures for the BFKL dynamics, since in such cases the DGLAP, or collinear configurations are suppressed. For the purpose of this analysis, we shall consider $\gamma^* \gamma^*$ scattering \cite{Brodsky:1996sg,Brodsky:1997sd}. Such process could be realized in the future $e^+e^- $ collider, such as FCC-ee \cite{Abada:2019lih,Abada:2019zxq}, or CLIC \cite{Linssen:2012hp}, with doubly tagged electrons in the final state. We compute this process using the BFKL resummed solution, and quantify the length of the preasymptotic regime. We find that the cross section which includes BFKL is in fact lower than the Born cross section (i.e. just a two gluon exchange) for large region of energy (this was also observed in \cite{Kwiecinski:1996td} which used BFKL with kinematical constraint) which increases with the increase of the scales of the virtual photons.   Still, we  find that  an enhancement over the Born cross section of the order of factor 2 to 3 can be attained for energies of $W\simeq 300 \, \rm GeV$ provided the scales of the virtual photons are less than $10 \, \rm GeV^2$. We also investigate the contributions of different transverse momenta to this process. We find that even though the impact factors are peaked at the scales of virtualities of the photons, there is significant spread in the transverse momenta along the exchanged Pomeron. This leads to the feature that the energy dependence of the resulting cross section is only weakly dependent on the virtualities of the photons.

The paper is organized as follows.  In the next section, Sec.~\ref{sec:bfkl} we introduce the LL and NLL BFKL formalism, and recall the sources of large corrections in the NLL BFKL using Mellin space representation. In Sec.~\ref{sec:ccss} we recall the construction of the CCSS resummation scheme. Following that,  in Sec.~\ref{sec:numerical_green}, we perform detailed numerical analysis of CCSS, focusing on the preasymptotic features of this solution. In particular, we quantify the dip or a plateau in the rapidity before the evolution starts to be dominated by the power-like growth typical of BFKL. In Sec.~\ref{sec:gammagamma}
we apply the resummed CCSS solution to the selected process of $\gamma^*\gamma^*$ scattering, and analyze the dependence on the energy, polarization, scales and investigate the dominant  transverse momenta which contribute to the BFKL Pomeron in this process. Finally, in last section we briefly summarize the results.

\section{LL and NLL BFKL evolution}
\label{sec:bfkl}
We shall consider BFKL evolution in the forward case $t=0$.
The BFKL \cite{Lipatov:1985uk} equation can be formulated as an evolution equation differential in rapidity and  it can be cast in the following generic form
\be
\frac{\partial}{\partial y} G(y;\kt,\kt_0) = \delta^2(\kt-\kt_0) + \int \frac{d^2 \kt'}{\pi} \;
\CK(\kt,\kt',\asbar) \; G(y;\kt',\kt_0) \; ,
\label{eq:bfkl_general}
\ee
where $G(y;\kt',\kt_0)$ is the forward BFKL gluon Green's function and $\CK(\kt,\kt')$ is the BFKL kernel which possesses the following perturbative expansion
\be
\CK(\kt,\kt',\asbar) = \asbar(\mu^2) \CK_{0}(\kt,\kt') + \asbar^2(\mu^2) \CK_{1}(\kt,\kt') + \dots \; ,
\label{eq:bfklkernel}
\ee
Here,  $\CK_{0}$ is leading logarithmic (LL) and $\CK_{1}$ is  next-to-leading logarithmic(NLL) BFKL kernel respectively.  In Eq.~\eqref{eq:bfkl_general} $y$ is the evolution variable, the rapidity, and  $\kt,\kt',\kt_0$ are transverse momenta of the reggeized gluons exchanged in the $t$ channel. In the following, we shall also use the  notation $|\kt^2|=k^2$.  The (rescaled) strong coupling constant is defined as
\be
\asbar(\mu^2) =  \frac{N_c}{\pi}\alpha_s(\mu^2) \;  ,
\ee
with $N_c$ the number of colors.  
The kernels $\CK_{0},\CK_{1}$ can be also transformed to Mellin space, by use of the following relation
\be
\chi(\gamma) \; = \;   \int_0^{\infty} dk'^2 \, \CK(k,k') \, \bigg(\frac{k'^2}{k^2}\bigg)^\gamma \; ,
\label{eq:mellin}
\ee
where we have used the angular averaged version of the kernel. In the Mellin space the BFKL kernel has the expansion
\be
\chi(\gamma)=\asbar \chi_0(\gamma) + {\asbar^2} \chi_1(\gamma)+ {{\cal O}(\asbar^3)} \; .
\ee
In QCD the kernels are known up to NLL order \cite{Fadin:1998py,Ciafaloni:1998gs}
and the  corresponding terms  at the  LL and NLL  level have 
 the following form in Mellin space
\be
\chi_{0}(\gamma) = 2\psi(1)-\psi(\gamma)-\psi(1-\gamma) \; ,
\label{eq:chi0}
\ee
and 
\begin{multline}
 \chi_1(\gamma) = -\frac{b}{2} [\chi^2_0(\gamma) + \chi'_0(\gamma)]
  -\frac{1}{4} \chi_0''(\gamma)
  -\frac{1}{4} \left(\frac{\pi}{\sin \pi \gamma} \right)^2
  \frac{\cos \pi \gamma}{3 (1-2\gamma)}
  \left(11+\frac{\gamma (1-\gamma )}{(1+2\gamma)(3-2\gamma)}\right) \\
 \quad +\left(\frac{67}{36}-\frac{\pi^2}{12} \right) \chi_0(\gamma)
  +\frac{3}{2} \zeta(3) + \frac{\pi^3}{4\sin \pi\gamma}   - \Phi(\gamma) \; ,
\label{eq:nllorg}
\end{multline}
with
\begin{equation}
\Phi(\gamma) = \sum_{n=0}^{\infty} (-1)^n
\left[ \frac{\psi(n+1+\gamma)-\psi(1)}{(n+\gamma)^2}
+\frac{\psi(n+2-\gamma)-\psi(1)}{(n+1-\gamma)^2} \right] \;,
\end{equation}
where $b=(33-2N_f)/(12\pi)$  with $N_f$ the number of (active) flavors and 
\begin{equation}
\psi(\gamma)=\frac{1}{\Gamma(\gamma)}\frac{d \Gamma(\gamma)}{d\gamma}\, ,
\label{eq:polygamma}
\end{equation}
 is the polygamma function.

In  N=4 sYM  theory the kernel $\chi$ is known up to NNLL order \cite{Velizhanin:2015xsa,Gromov:2015vua,caron2018high}.
It is also worth pointing out that in QCD the functions $(k'^2)^\gamma$ are the eigenfunctions of the equation, and $\chi_0(\gamma)$ is the eigenvalue at the leading logarithmic order. Strictly speaking this is not the case in the NLL order in QCD due to the fact that the coupling constant is running and the conformal invariance is broken. In that case,
the eigenfunctions are different, and include the running of the coupling, see \cite{Chirilli:2013kca}. This however does not preclude the possibility of performing Mellin transform of the BFKL equation as in Eq.\eqref{eq:mellin} and analyzing the Mellin transform of the NLL part of the kernel in the form \eqref{eq:nllorg}. 

The above kernel in NLL has been written for the so-called symmetric choice of scales. As discussed in \cite{Fadin:1998py,Ciafaloni:1998gs,Salam:1998tj} at NLL level  the scale choice does matter. Let us briefly recall the scale changing transformations. In general, the  generic cross section for two scale process in the high energy factorization  can be written in the following form

\begin{equation}
\sigma = \int \frac{d\omega}{2\pi i}  \int \frac{d^2 \kt}{\kt^2} \frac{d^2 \kt_0}{\kt_0^2} \left(\frac{s}{Q Q_0}\right)^{\omega} h^{A}(Q,\kt) G_{\omega} (\kt,\kt_0) h^{B}(Q_0,\kt_0) \;.
\end{equation}

Here $h^{A}(Q,\kt),h^{B}(Q_0,\kt_0)$ are the impact factors, for example for the jet production in the case of the Mueller-Navelet process \cite{Colferai:2010wu} or as will be studied in this paper for the photon-gluon impact factors.
In the first case $Q,Q_0$ would correspond to the transverse momenta of the hard jets or to the virtualities of the photons in the $\gamma^*\gamma^*$ scattering. The relation for the Green's function in Mellin space and in rapidity space in this case is equal to 
\begin{equation}
G(y;\kt,\kt_0) = \int \frac{d\omega}{2\pi i} \, e^{\omega y} \, {\cal G}_{\omega}(\kt,\kt_0)\; ,
\end{equation}
where evolution variable for the BFKL is defined as $y= \ln \frac{\nu}{ k k_{0}}$, and $\nu$ is the energy available for the BFKL evolution.  In this example the energy scale choice is symmetric, but it might as well be asymmetric $\nu/k^2$ or $\nu/k_{0}^2$.  This will imply the change of the gluon Green's function as well
\begin{equation}
G_{\omega} \longrightarrow G_{\omega} \left(\frac{k_{>}}{k_{<}}\right)^{\omega}\,, \quad k_> = \max(k,k_0), \quad k_<=\min(k,k_0)\; ,
\end{equation}
and in consequence there are similarity transformations \cite{Ciafaloni:1998gs,Ciafaloni:2003rd} for the 
BFKL kernel

\begin{equation}
K_{\omega}^{s_0=k^2}(k,k_{0}) = K_{\omega}^{s_0=k k_{0}}(k,k_{0}) \left( \frac{k}{k_{0}} \right)^{\omega}, \quad \quad K_{\omega}^{s_0=k_0^2}(k,k_{0}) = K_{\omega}^{s_0=k k_{0}}(k,k_{0}) \left( \frac{k_0}{k} \right)^{\omega} \, .
\end{equation}

The expression Eq.~\eqref{eq:nllorg} is given for the symmetric scale choice. In order to transform to the asymmetric scale choice, for example for $s_0=k^2$ in the Mellin space one needs to add to  Eq.~\eqref{eq:nllorg}, the term
\begin{equation}
-\frac{1}{2} \chi_0(\gamma) \frac{\partial \chi_0}{\partial \gamma}\;,
\end{equation}
which is the NLL term due to the scale changing transformation.  At NNLL the scale changing transformation is much more complicated as it  involves the NLL kernel as well  \cite{Marzani:2007gk,Deak:2019wms}.

Coming back to the NLL kernel, the well known problem that arises at NLL order in BFKL is due to the presence of double and triple collinear logarithms. The collinear logarithms manifest themselves as poles in the $\gamma$ plane. The collinear poles, which correspond to the strong ordering when $k^2 \gg k'^2$ appear as $1/\gamma$ and the anti-collinear poles which correspond to $k^2 \ll k'^2$ as $1/(1-\gamma)$. The double poles are arising due to the running coupling and the non-singular  (in $1/z$) part of the DGLAP splitting function which appear at NLL order. To be precise, keeping most singular $\gamma \rightarrow 0$ and $\gamma \rightarrow 1$ contributions gives for the corresponding terms
\be
-\frac{b}{2} [\chi^2_0(\gamma) + \chi'_0(\gamma)] \;\;\; \longrightarrow \;\;\; -b \frac{1}{(1-\gamma)^2} \; ,
\ee 
for the running coupling,
\be
 \frac{\cos \pi \gamma}{3 (1-2\gamma)}
  \left(11+\frac{\gamma (1-\gamma )}{(1+2\gamma)(3-2\gamma)}\right) \;\;\; \longrightarrow \;\;\;  -\frac{11}{12} \frac{1}{\gamma^2}\, -\, \frac{11}{12} \frac{1}{(1-\gamma)^2} \; ,
\ee
for the DGLAP contribution. There are also triple collinear poles which appear due to the kinematical constraints \cite{Ciafaloni:2003ek,Ciafaloni:2003rd}. Such constraints were discussed in the BFKL context as originating from the improved kinematics, and more precisely by the requirement that the exchanged momenta are dominated by the transverse components \cite{Andersson:1995ju,Kwiecinski:1996td}, for more recent work on different forms of kinematical constraint see \cite{Deak:2019wms}. These  contributions generate the double transverse logarithms and in the Mellin space they exhibit most singular behavior, the triple collinear poles
\be
-\frac{1}{4} \chi_0''(\gamma) \;\;\; \longrightarrow \;\;\; -\frac{1}{2} \frac{1}{\gamma^3}\, -\, \frac{1}{2} \frac{1}{(1-\gamma)^3}  \; .
\ee
As it has been demonstrated in \cite{Salam:1998tj,Salam:1999cn} this collinear approximation 
\be
\chi(\gamma)_{\rm coll} = -b \frac{1}{(1-\gamma)^2} -\frac{11}{12} \frac{1}{\gamma^2}\, -\, \frac{11}{12} \frac{1}{(1-\gamma)^2}-\frac{1}{2} \frac{1}{\gamma^3}\, -\, \frac{1}{2} \frac{1}{(1-\gamma)^3} \; ,
\ee
accounts for the major part of the NLL corrections given by $\chi_1$.
\section{Resummed CCSS scheme}
\label{sec:ccss}
\subsection{General setup}

Let us briefly remind the reader about  the content of  CCSS resummation scheme. General setup for the CCSS scheme was based on the analysis of poles in the Mellin space, but the final formulation and the solution to the equation was given in the momentum space.  A similar  idea for the resummation was formulated previously also in Ref.~\cite{Kwiecinski:1997ee} by combining the DGLAP and BFKL evolution with the kinematical constraint.  In the  CCSS resummation one subtracts triple and double poles and incorporates  the full DGLAP splitting function and the kinematical constraint which both resum double and triple poles respectively. 
In addition, more subtractions are needed to ensure the conservation of the momentum sum rule.

In the original  CCSS scheme \cite{Ciafaloni:2003ek,Ciafaloni:2003rd} one starts with the LL+NLL BFKL kernel  with LO DGLAP splitting function and imposes kinematical constraint. The double and triple poles are then subtracted in order to avoid the double counting. Also, the running coupling poles are subtracted from the NLL expression since the running coupling is later on incorporated directly in front of the LL kernel with the scale $\qt$, which is the transverse momentum of the emitted gluon. The expression for the resummed kernel NLL part of the kernel in the Mellin space is then
\begin{equation}
\chi_{\rm resum}^1(\gamma) \; = \;  \chi_1(\gamma)
+\frac{1}{2} \chi_0(\gamma)  \frac{\pi^2}{\sin^2(\pi\gamma)}-\chi_0(\gamma) \frac{A_1(0)}{\gamma(1-\gamma)}+\frac{b}{2} (\chi_0'+\chi_0^2) \;.
\label{eq:subtractions}
\end{equation}
Of course one could change the scale choice in the running coupling in front of the LL kernel, which would change the term at NLL level. The choice of $q^2$ is such that no other $b$ dependent terms are present in the LL kernel. The choice of the scale for the coupling constant in the NLL kernel is arbitrary in the sense that any change is of NNLL level which is beyond the control. In CCSS scheme the  scale at NLL was chosen to be the maximum of  the momenta squared of the exchanged gluons, i.e.
$\max(k^2,k^{\prime 2})$.
\subsection{Momentum representation}
The Mellin space expression for the CCSS resummation was then transformed back to the momentum space, and the resummed evolution equation was directly solved in the momentum space. This was done so that running coupling could be incorporated more easily in   the evolution. With the DGLAP and kinematical constraint included, the equation becomes an integral equation in both longitudinal and transverse momentum ($z,\kt$).
The final resummed kernel $\CK(z;k,k')$ in the CCSS  method \cite{Ciafaloni:2003rd,Ciafaloni:2003ek} is the sum of three contributions:
\begin{multline}
 \int_x^1\frac{dz}{z}\int d k'^2 \; \CK(z;k,k') f(\frac{x}{z},k')\\
 = \int_x^1\frac{dz}{z}\int d k'^2
 \left[\bar{\alpha}_s(\qt^2) K_0^{\kc}(z;\kt,\kt') +
 \bar{\alpha}_s(k_{>}^2) K_{\ci}^{\kc}(z;k,k')+
 \bar{\alpha}^2_s({k}^2_{>}) \tilde{K}_1(k,k') \right ] f(\frac{x}{z},k') \;.
 \label{eq:kernelxk}
\end{multline}
The first term in Eq.~\eqref{eq:kernelxk} is the  LL BFKL with running coupling and consistency constraint imposed onto real part of the kernel
\begin{multline}
\label{eq:LOBFKLkc}
 \int_x^1\frac{dz}{z}\int dk'^2 \; \left[ \bar{\alpha}_s(\qt^2)
  K_0^{\kc}(z;\kt,\kt') \right] f(\frac{x}{z},k') \\
 = \int_x^1 \frac{dz}{z} \int {d^2 \qt \over \pi \qt^2} \;
  \bar{\alpha}_s(\qt^2) \left[ f(\frac{x}{z},|\kt+\qt|)
 \Theta(\frac{k}{z}-k')\Theta(k'-kz)-\Theta(k-q) f(\frac{x}{z},k) \right]\;,
\end{multline}
where $\qt = \kt-\kt'$.
The consistency or kinematical constraint appears in the form of the $ \Theta(\frac{k}{z}-k')\Theta(k'-kz)$ functions, and is here symmetric, which corresponds to the symmetric scale choice.  To be precise, there are different versions of this constraint which appear in the literature, see \cite{Andersson:1995ju,Kwiecinski:1996td,Ciafaloni:1987ur}. It turns out that all versions are generating the same leading cubic poles in the Mellin space at NLL, they generate no double poles, and with the difference starting to appear in the single pole level, see \cite{Deak:2019wms}. The scale in the  running coupling  corresponds to the transverse momentum  of the emitted gluon $\qt^2$.
The second term in \eqref{eq:kernelxk} is the non-singular (in $z$ ) DGLAP splitting function terms with consistency constraint
\begin{multline}
 \int_x^1\frac{dz}{z}\int d k'{}^2 \; \bar{\alpha}_s(k_{>}^2)
  K_{\ci}^{\kc}(z;k,k') f(\frac{x}{z},k') =  \\
 = \int_x^1 { dz \over z} \int_{(kz)^2}^{k^2} \frac{{dk'}^2}{k^2} \;
  \bar{\alpha}_s(k^2) z{k \over k'} \tilde{P}_{gg}(z{k \over k'})
  f(\frac{x}{z},k')   
 + \int_x^1 { dz \over z} \int_{k^2}^{(k/z)^2} \frac{{dk'}^2}{k'{}^2} \;
  \bar{\alpha}_s({k'}^2) z{k' \over k} \tilde{P}_{gg}(z{k' \over k})
  f(\frac{x}{z},k') \;,  
  \label{eq:dglapterms}
\end{multline}
Finally, the last  term is the  NLL part of the BFKL with subtractions given by Eq.~\eqref{eq:subtractions}, all transformed into momentum space
\begin{align}
 \int_x^1\frac{dz}{z} & \int d k'{}^2 \;   \bar{\alpha}^2_s({k}^2_{>})
  \tilde{K}_1(k,k') f(\frac{x}{z},k')  \; = \;  
   {1\over 4}\int_x^1 \frac{dz}{z} \int d{k'}^2 \; \bar{\alpha}^2_s({k}^2_{>})
  \bigg\{  \\
& {\left({67 \over 9} - {\pi^2 \over 3}\right) {1\over |{k'}^2-k^2|}
  \left [f(\frac{x}{z},{k'}^2) - {2 k_{<}^2 \over ({k'}^2 + k^2)}
  f(\frac{x}{z},k^2)\right] + } \\
& {\bigg[ - {1 \over 32} \left({2 \over {k'}^2} + {2 \over k^2} +
  \left({ 1\over {k'}^2 } - {1\over k^2} \right)
  \log\left({k^2 \over {k'}^2}\right)\right)
  + {4 \Li(1-k_{<}^2/k_{>}^2) \over |{k'}^2 - k^2|}} \\
 &{-4 A_1(0){\rm sgn}({k}^2-{k'}^2)
  \left( {1 \over k^2} \log{|{k'}^2-k^2| \over {k'}^2} -
  {1 \over {k'}^2} \log{|{k'}^2-k^2| \over {k}^2}\right)} \\
& - \left(3 + \left({3 \over 4} - {({k'}^2+k^2)^2 \over 32{k'}^2 k^2}\right)
  \right) \int_0^{\infty} {dy \over k^2 + y^2 {k'}^2 }
  \log|{1+y \over 1-y}| \\
 &+ {1 \over {k'}^2 + k^2} \left( {\pi^2 \over 3} +
  4 \Li( {k_{<}^2 \over k_{>}^2})\right) \bigg]
  f(\frac{x}{z},k') \bigg\} \\
& + {1\over 4} 6 \zeta(3) \int_x^1 \frac{dz}{z} \;
  \bar{\alpha}^2_s(k^2)  f(\frac{x}{z},k) \;. \hspace{40mm}
\end{align}

The non-singular (as $z\rightarrow 0$) splitting function in the DGLAP terms appearing in Eq.~\eqref{eq:dglapterms} is defined as follows:
\begin{equation}
 \tilde{P}^{(0)}_{gg}  =  P^{(0)}_{gg} - { 1 \over z} \;,
\end{equation}
where the $ P^{(0)}_{gg}$ is the   DGLAP gluon-gluon splitting function in LO
(in this work we only consider purely gluonic channel, $N_f=0$).

Following arguments presented in \cite{Ciafaloni:2003rd}  we note that the argument of the splitting function ${\tilde P}$ has to be
shifted in Eq.~\eqref{eq:dglapterms} in order to reproduce the correct collinear
limit when the kinematic constraint ($kz < k' < \frac{k}{z}$) is included.

In addition to  the terms presented above, one needs also to include the terms which will cancel spurious DGLAP anomalous dimension at NLO. This was done by adding extra terms. Since there is an ambiguity in this  procedure, two different schemes were proposed $A$,$B$ in \cite{Ciafaloni:2003ek}. In here, we shall utilize scheme $B$, which  importantly also satisfies the momentum sum rule.

\section{Numerical results for the gluon Green's function}
\label{sec:numerical_green}
\begin{figure}
	\begin{center}
		\includegraphics[width=0.49\textwidth]{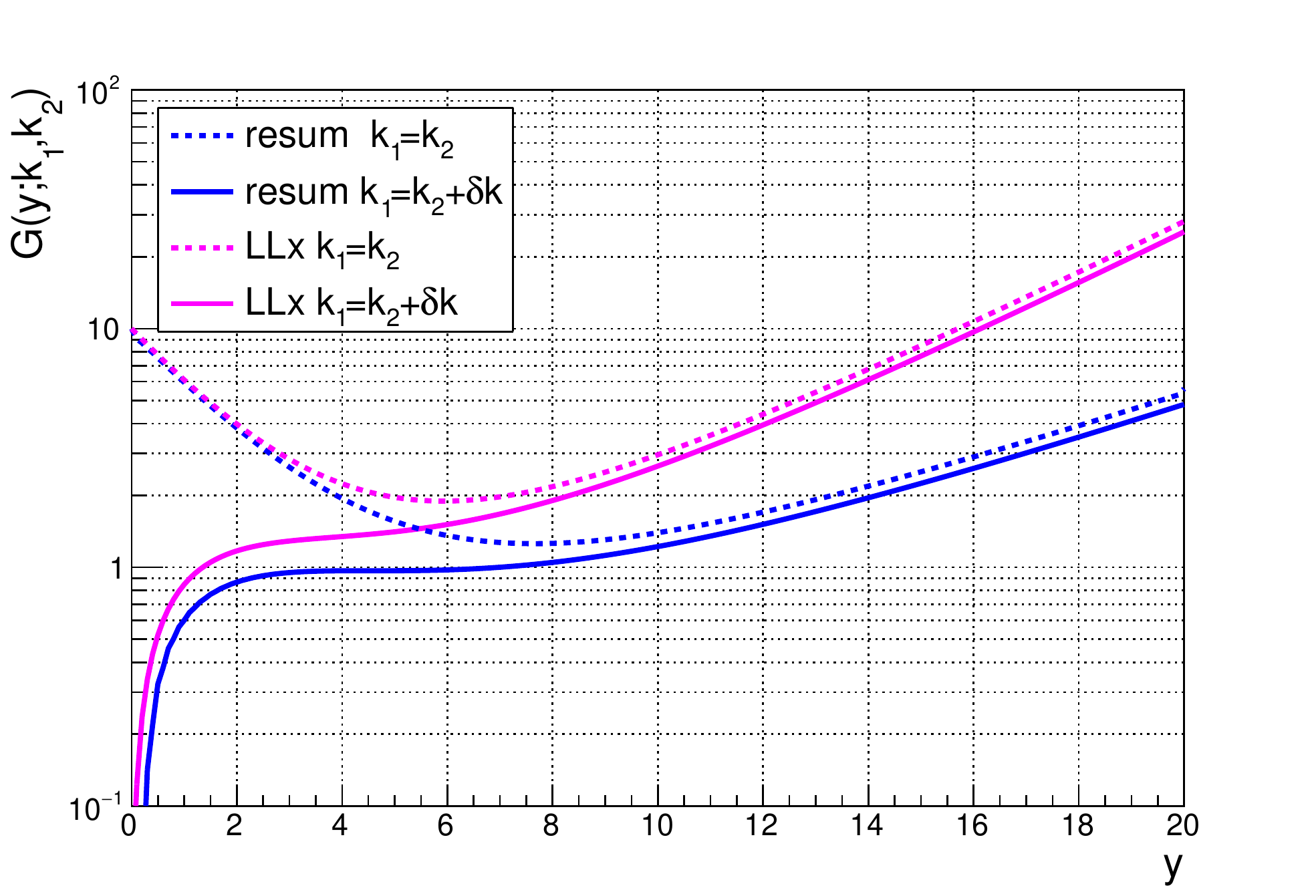}
		\includegraphics[width=0.49\textwidth]{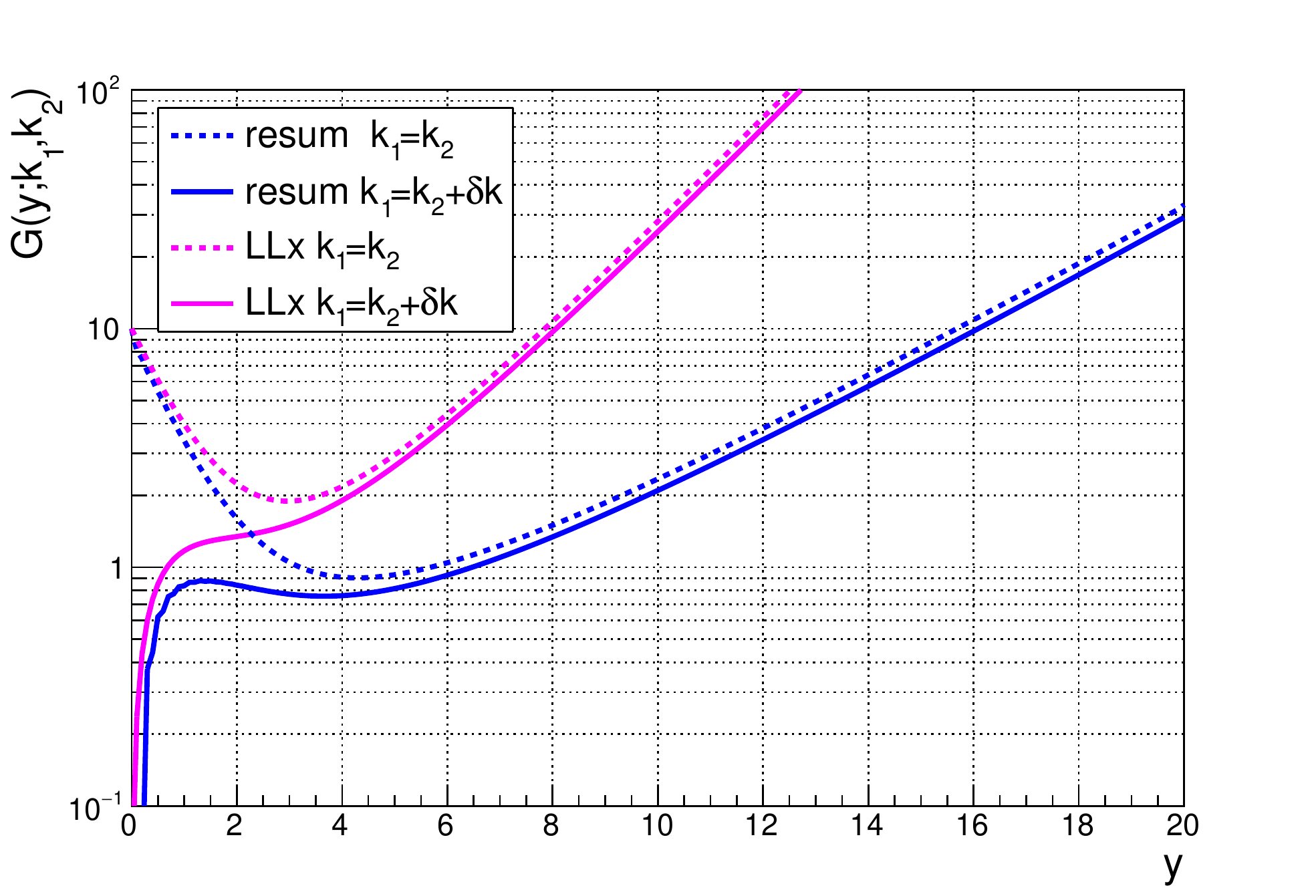}
		\caption{Gluon Green's function $G(y;k_1,k_2)$ as a function of rapidity $y$ for fixed transverse momentum $k_1 = 5 \, {\rm GeV}$ and fixed coupling constant $\asbar=0.1$ (left plot) and
			$\asbar=0.2$ (right plot). LLx (magenta) and resummed (blue) calculations are shown for comparison.  Dashed lines correspond to $k_1=k_2$, and solid lines $k_1=k_2+\delta k$ with $\delta k $ an offset which is given by $\delta k= k_1 (\exp(\delta \ln k)-1)$ where $\delta \ln k$ is the spacing in the logarithmic grid used in the numerical solution.}
	\end{center}
	\label{fig:lovsresum1}
\end{figure}

We start the presentation of the numerical results for the gluon Green's function $G(y;k_1,k_2)$ which is the solution to the resummed BFKL evolution equation.
In this analysis we shall focus on the preasymptotic features of the resummed solution and try to quantify them. The numerical method for the solution of this equation was identical to the one used in \cite{Ciafaloni:2003rd}. We start from the fixed coupling case, and compare the solutions of the leading order (LLx) BFKL with the resummed evolution. The solutions are demonstrated in Fig.~\ref{fig:lovsresum1}, for two values of the fixed coupling $\asbar=0.1, 0.2$. 
Two sets of curves are presented, one for the resummed case (blue) and one for the LL case (magenta). In each case, we show the gluon Green's function $G(y;k_1,k_2)$ and $G(y;k_1,k_1+\delta k)$, where $\delta k$ is a small value. In our case it is equal to the distance between the points on the logarithmic grid in $\ln k$, i.e. $\delta k=k_1 (\exp(\delta \ln k)-1)$ where $\delta \ln k$ is the spacing in the grid. The reason for that, as explained in \cite{Ciafaloni:2003rd}, is that for exactly equal values of scales the preasymptotic behavior is dominated by the numerical `delta' function, and thus leads to a minimum that partially depends on the numerical details, like the grid spacing. This is evident when we compare the solutions for  $G(y;k_1,k_1)$  and 
$G(y;k_1,k_1+\delta k)$.
We observe that as expected, the resummed evolution exhibits much slower rise than the LLx evolution which is related to the lower value of the resummed intercept.  However, this is not the only reason that the resummed solution leads to much smaller numerical value for the Green's function. On top of that   there are substantial preasymptotic effects, which manifest themselves as the presence of the `plateau'  or even a  minimum in the resummed case. Such `plateau' is  almost completely absent in the LLx case, though the preasymptotic effects are present also at this order and are visible through the slower growth in the first few units of rapidity. We observe that the width of the flat region  in rapidity in resummed case strongly depends on the value of the strong coupling, and it can extend for many units of rapidity. For values of the coupling of the order of $\bar{\alpha}_s =  0.1$, the `plateau' extends to about 8 units of rapidity and in the case of $\bar{\alpha}_s=0.2$ it is about 4-5 units of rapidity.
\begin{figure}
	\begin{center}
		\includegraphics[width=10cm]{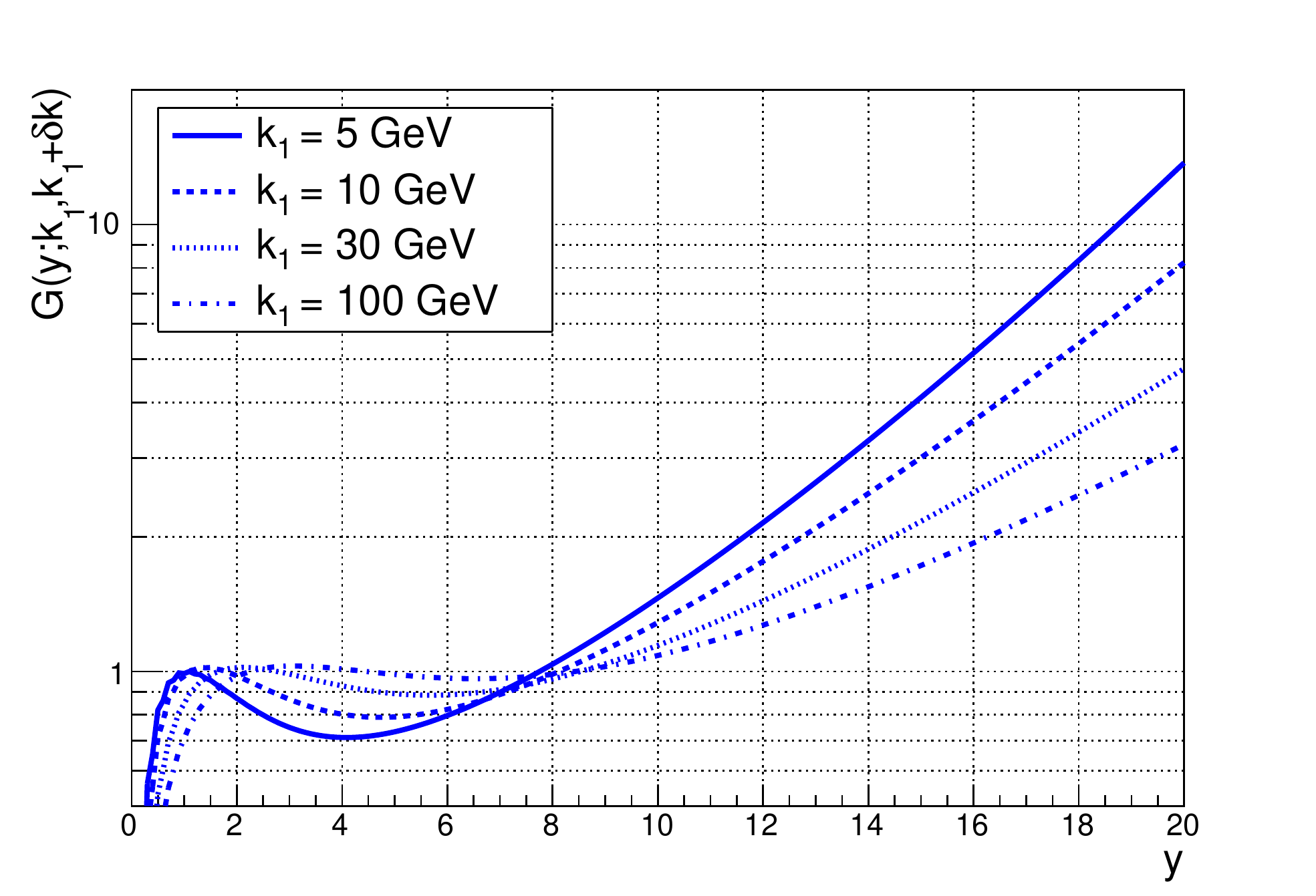}
		\label{fig:resumrccomp}
		\caption{Gluon Green's function $G(y;k_1,k_2)$ as a function of rapidity $y$ for fixed transverse momenta $k_1 = 5,10,30,100 \, {\rm GeV}$ and running strong coupling constant. }
		\label{fig:bfklresum_rc}
	\end{center}
\end{figure}

In Fig.~\ref{fig:bfklresum_rc} we plot the gluon Green's function as a function of rapidity for the running coupling case. We observe similar feature in the case of the calculation with the running coupling included. In this case the position of the dip or a preasymptotic plateau and the onset of the increase depends on the value of the scales for which the gluon Green's function is evaluated. In general for larger scales, the onset of the increase is delayed to larger rapidities.

\begin{figure}
	\centerline{\includegraphics[width=16cm]{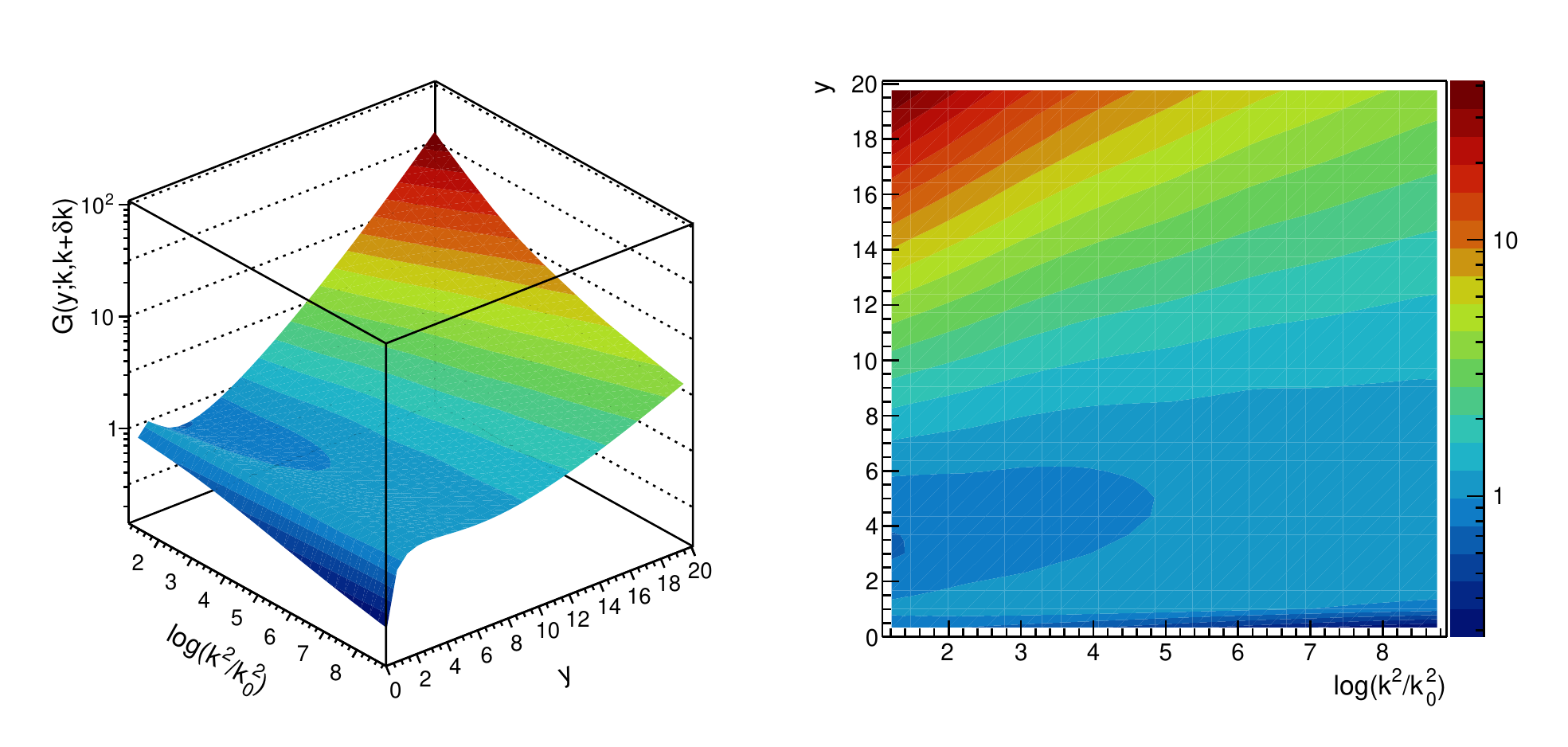}}
	\caption{Gluon Green's function as a two-dimensional function of $\log k$ and $y$. The scale is set to be $k_0=1 \, \rm GeV$.}
	\label{fig:green_2d_resum}
\end{figure}

The preasymtotic plateau can be  better illustrated in the two dimensional plot Fig.~\ref{fig:green_2d_resum} where we show $G(y;k_1,k_1+\delta k)$ as a two-dimensional surface,
as well as contour plot in $(y,\log(k_1))$ space.  The `dip' in rapidity is most prominent in the low $k^2$ region, and it is clear that for several units of rapidity the growth is very slow.

In order to quantify the dependence of the preasymptotic region on the external scales, we take the solution with equal scales and evaluate the minimum of this function. As can be seen from Fig.~\ref{fig:lovsresum1} the position of the minimum in this case gives a good estimate of the onset of the rise in rapidity. 

\begin{figure}
	\centerline{\includegraphics[width=10cm]{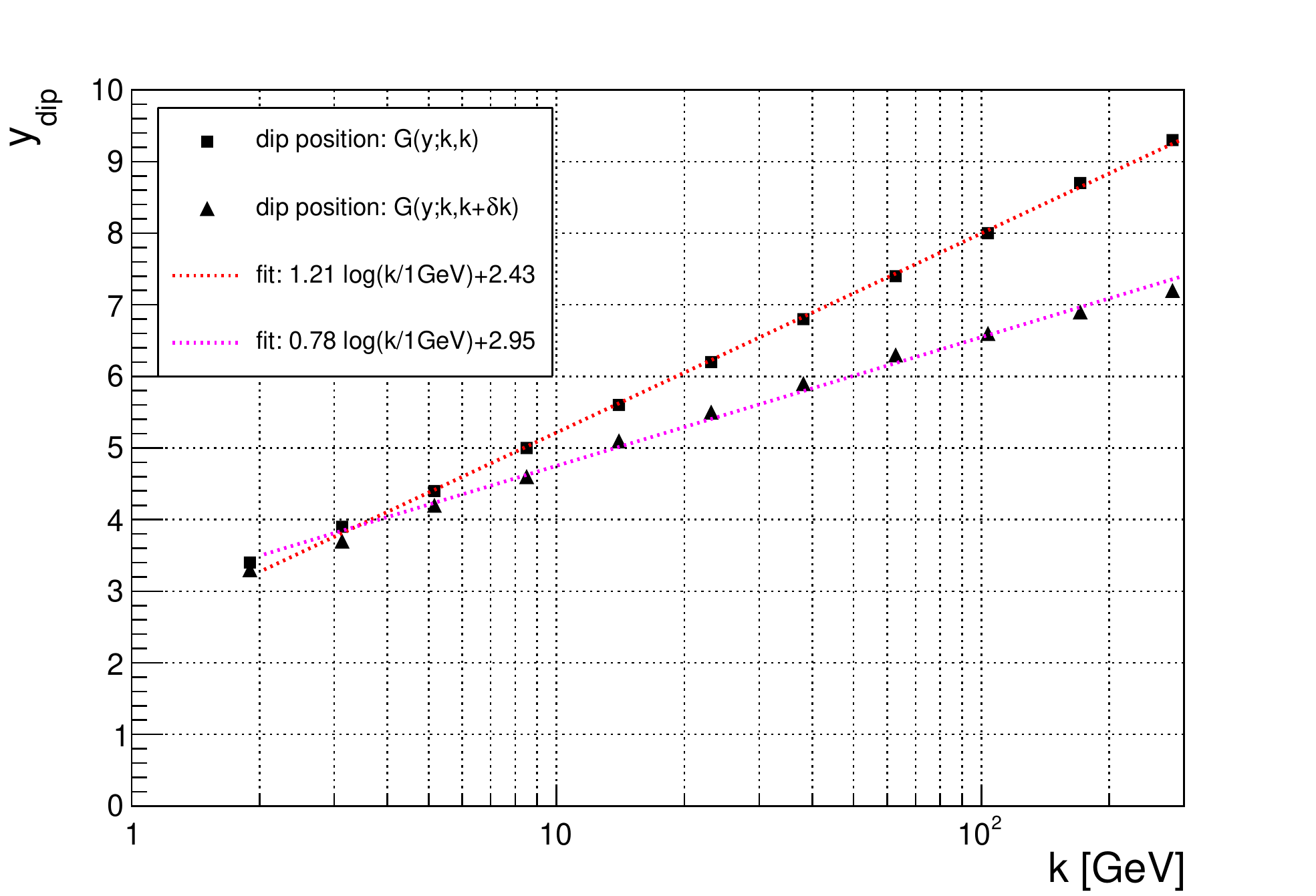}}
	\label{fig:dipkt}
	\caption{Dip position in rapidity of the resummed gluon Green's function $G(y;k,k)$ (squares)
		and $G(y;k,k+\delta k)$ (triangles) as a function of the value of  the scale $k$. Strong coupling is running. The lines correspond to the fits.}
\end{figure}

\begin{figure}
	\centerline{\includegraphics[width=10cm]{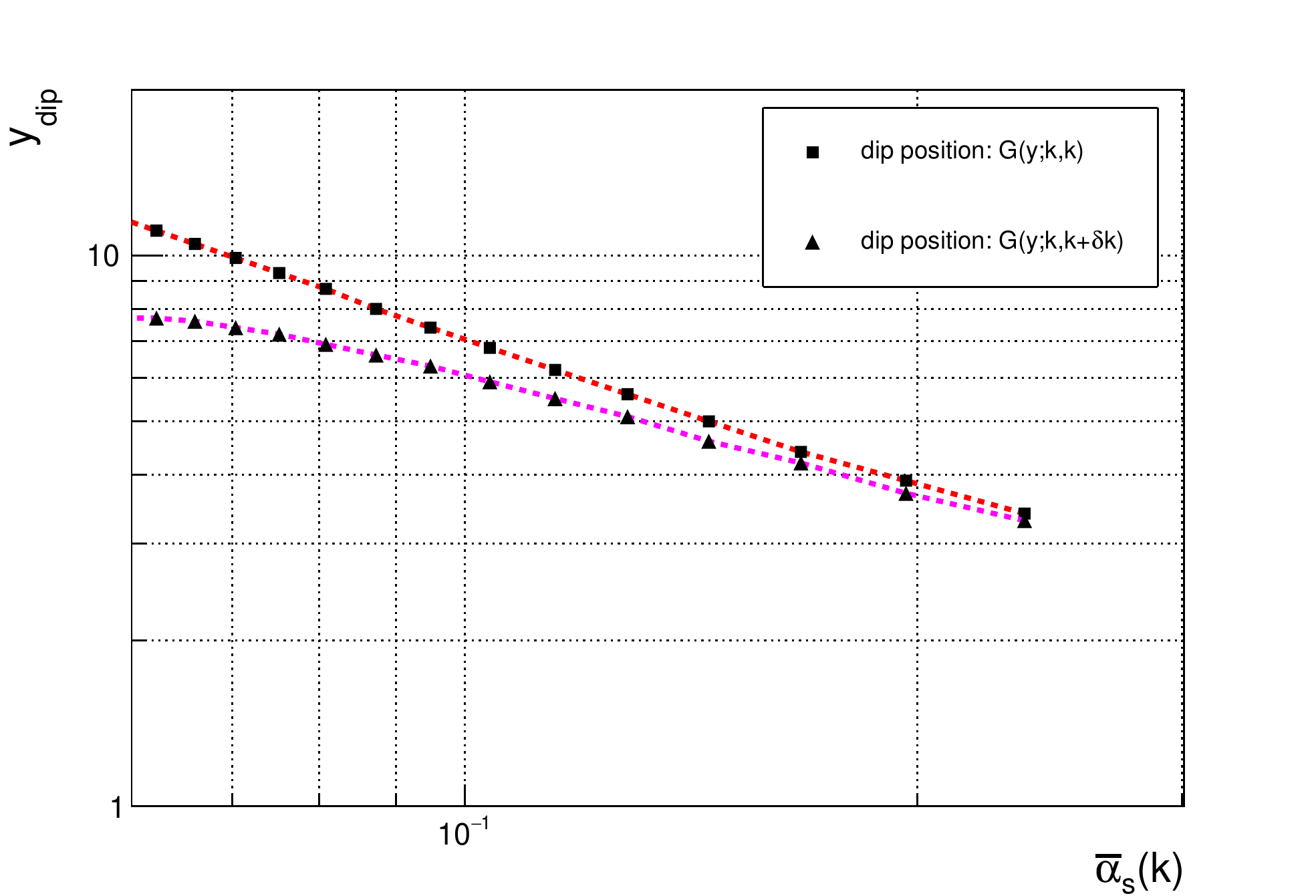}}
	\label{fig:dipas}
	\caption{Dip position in rapidity of the resummed gluon Green's function $G(y;k,k)$ (squares)
		and $G(y;k,k+\delta k)$ (triangles) as a function of the value of strong coupling constant $\bar{\alpha}_s$ evaluated at the scale $k$. Dashed lines are connecting the points and are just to guide the eye.}
\end{figure}

The position of the minimum in rapidity as a function of the scale in the gluon Green's function is shown in Fig.~\ref{fig:dipkt}.  We performed two calculations, where we measure the position of the minimum in $G(y,k,k)$ and in $G(y,k,k+\delta k)$. Calculations in Fig.~\ref{fig:dipkt} were done with running strong coupling. We see that to a very good approximation the dependence on the transverse momentum is logarithmic. In fact the linear fit shown in this figure describes the extracted points very well. This suggests, that the dependence on the value of the strong coupling is like $\sim 1/\bar{\alpha}_s(k)$.  In fact the linear fit works very well for the case of the dip in $G(y;k,k)$  and slightly worse for the $G(y;k,k+\delta k)$. In the latter case there  is some curvature, visible especially  when going to higher values of $k$. 
In any case, the position of the dip varies from  about 3 units of rapidity for $k=2 \, \rm GeV$ to about  $7-9$ for $k=300 \, \rm GeV$. This indicates that the preasymptotic effects are rather large and significantly delay the onset of the BFKL regime with Pomeron-like growth.
The approximately linear dependence on   $\sim 1/\bar{\alpha}_s(k)$ is  demonstrated  in the plot shown in Fig.~\ref{fig:dipas}, which shows straight line in the double logarithmic axis for the dip in the function $G(y;k,k)$. The slight curvature in the case of $G(y;k,k+\delta k)$ is also visible. Such simple dependence on the value of the coupling constant suggests that the minimum occurs for particular values of the product $y_{\rm dip} \alpha_s$. The value of this parameter $y_{\rm dip} \alpha_s$ is approximately equal to $0.7-0.8$ and it is almost constant when varying value of $k$.

The inverse relation of the position of the minimum as a function of the strong coupling could be expected from the analytic form of the solution to the gluon Green's function. It is well known that the saddle point approximation leads to the form
\begin{equation}
G(Y,k_\perp,k_\perp^\prime) \sim \frac{1}{\sqrt{Y}} e^{\omega Y} \exp(-\frac{\log({k_\perp}{k_\perp^{\prime}}/k_0^2)}{D Y}) \; ,
\end{equation}
which leads to the minimum at
\begin{equation}
\omega Y = \frac{1}{2} \; ,
\end{equation}
Thus one expects the minimum to occur even at LL order, for the values of $\asbar Y \sim 0.2$ and in that case it indeed should go like $1/\asbar$.

Thus from this numerical analysis, we see that the LL and resummed solutions vary not only with respect to the speed of the evolution, but also with respect to where exactly the BFKL evolution sets in. This can lead to much greater differences numerically than one would naively expect if the difference was only in  the intercept.

\begin{figure}
	\centerline{
	\includegraphics[width=10cm]{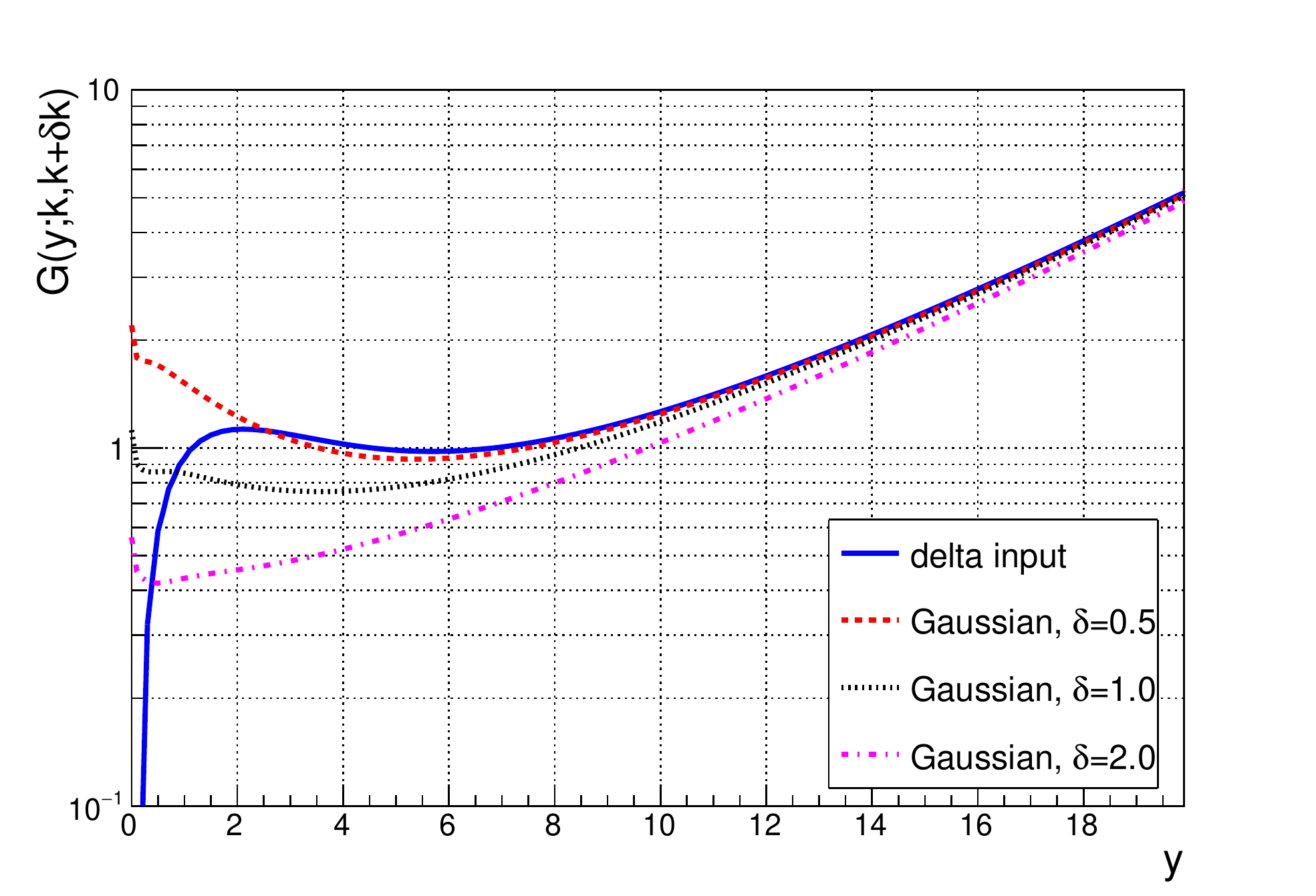}}
	\label{fig:inputvariation}
	\caption{The gluon Green's function $G(y;k,k+\delta k)$ as a function of rapidity for $k=30\,\rm GeV$. Three different initial conditions are considered: delta function (blue solid), and Gaussian initial conditions with widths equal to $\delta=0.5,1.0,2.0$ (red dashed, black dotted, magenta dashed-dotted) respectively. Initial conditions have been normalized.}
\end{figure}

The preasymptotic features will be however modified when convoluted with the impact factors due to the initial spread of the transverse momenta. Before we analyze it in the context of the physical process, we shall test the dependence on the initial conditions, by simply modifying the initial condition to be a Gaussian in $\log k^2$ with some width given by parameter $\delta$, i.e.
$G^{0} \sim \exp(-\log^2 (k^2/k_j^2)/\delta^2)$. 
In Fig.~\ref{fig:inputvariation} we show the solution for $k_j=30 \, \rm GeV$ for four cases: delta input, and three Gaussian inputs with different widths: $\delta=0.5,1.0,2.0$. The inputs have been normalized to unity, when integrated over $\log k$.
Interestingly all the solutions converge at high rapidity, and evolve with the same power asymptotically as they should.  The  dip and the slow down of evolution at initial rapidity is most prominent in the case of the delta function, and becomes less visible for broader initial conditions, i.e. when $\delta$ is increased.  On the other hand the broader initial conditions lead to solutions that initially start to increase faster with rapidity. Overall, we see that the dependence on the initial conditions is washed out only for rapidities of about 8 units, at which point the solutions converge.

\begin{figure}
	\centerline{
		\includegraphics[width=0.49\textwidth]{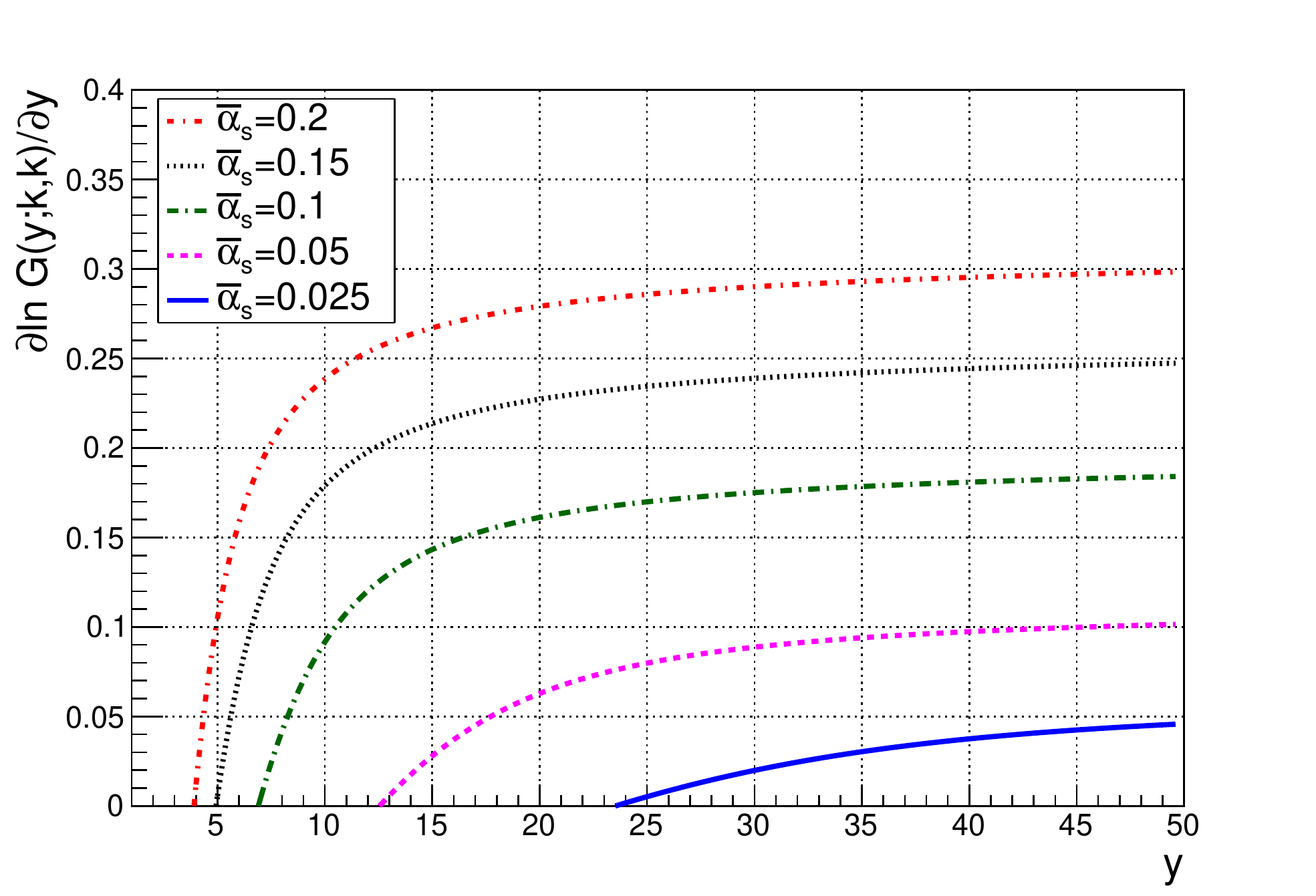}
		\includegraphics[width=0.5\textwidth]{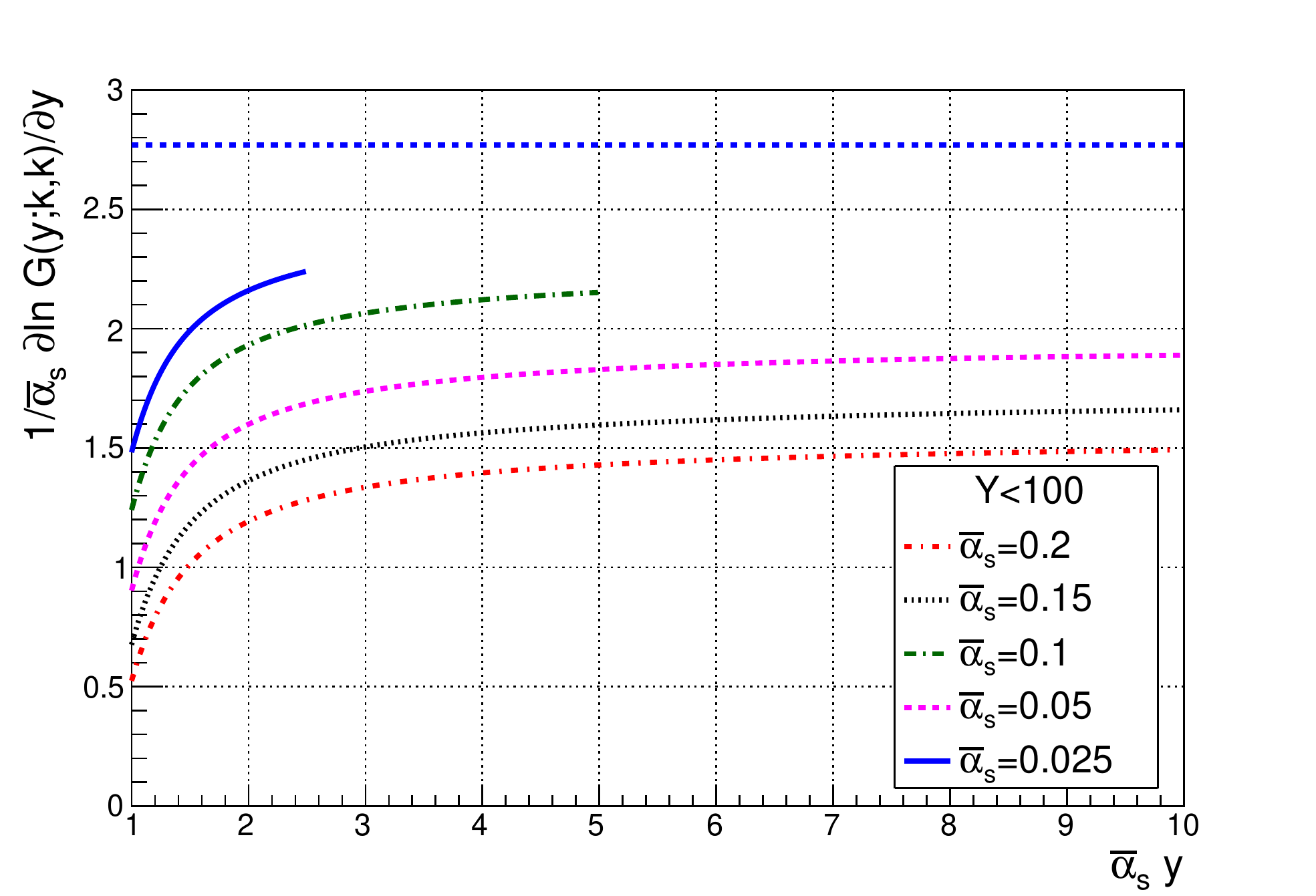}
	}
	\caption{Left plot: logarithmic derivative of the resummed gluon Green's function with respect to the  rapidity $y$ for fixed values of the strong coupling constant $\bar{\alpha}_s$ and for equal choice of the transverse momenta $k$. Right plot: the same quantity but rescaled by the inverse of the strong coupling constant and plotted as a function of the combined variables $\bar{\alpha}_s y$. Horizontal blue-dashed  line indicates LL value $4\ln2$. The reason that the curves for small values of the coupling constant are cutoff is that in all cases the evolution is run up to $Y<100$.}
	\label{fig:der_resum}
\end{figure}
As a last part in this section we analyze the approach to asymptotia by investigating the logarithmic derivative of the gluon Green's function with respect to the rapidity. To simplify the analysis we study here the fixed coupling case with all other corrections included.
In Fig.~\ref{fig:der_resum} (left plot) we show the logarithmic derivative of the gluon Green's function as a function of rapidity $y$. We see that one needs to evolve to very high rapidities for the derivative to go to a limiting values  for the intercept. For example, for value of $\bar{\alpha}_s=0.15$, one is close to asymptotic value of the intercept for rapidities larger than $20$.
The lower the values of $\alpha_s$, the higher the rapidity at which the asymptotic behavior sets in. The same quantity is plotted in a different way in Fig.~\ref{fig:der_resum}, as a function of combined variable $\bar{\alpha}_s y$ and it is rescaled by $1/\bar{\alpha}_S$. In the case of the leading logarithmic result this quantity should go to $\chi_0(\frac{1}{2})=4\ln2$ which is indicated as a dashed horizontal line. We see first that one needs to go to rather high values of $\bar{\alpha}_sy$ in order for the intercept to settle at an asymptotic value, and moreover these asymptotic values are still  away from the leading logarithmic value even at the smallest values of  the strong coupling $\sim 0.025$. Also, in that case, even for the highest values of rapidity the obtained value is much lower than the LLx result. That indicates, that in practice, for realistic energies one cannot expect the leading logarithmic result to be valid, even in approximate form.
\section{$\gamma^* \gamma^*$ scattering at high energy}
\label{sec:gammagamma}

\begin{figure}
	\begin{center}
		\includegraphics[width=5cm]{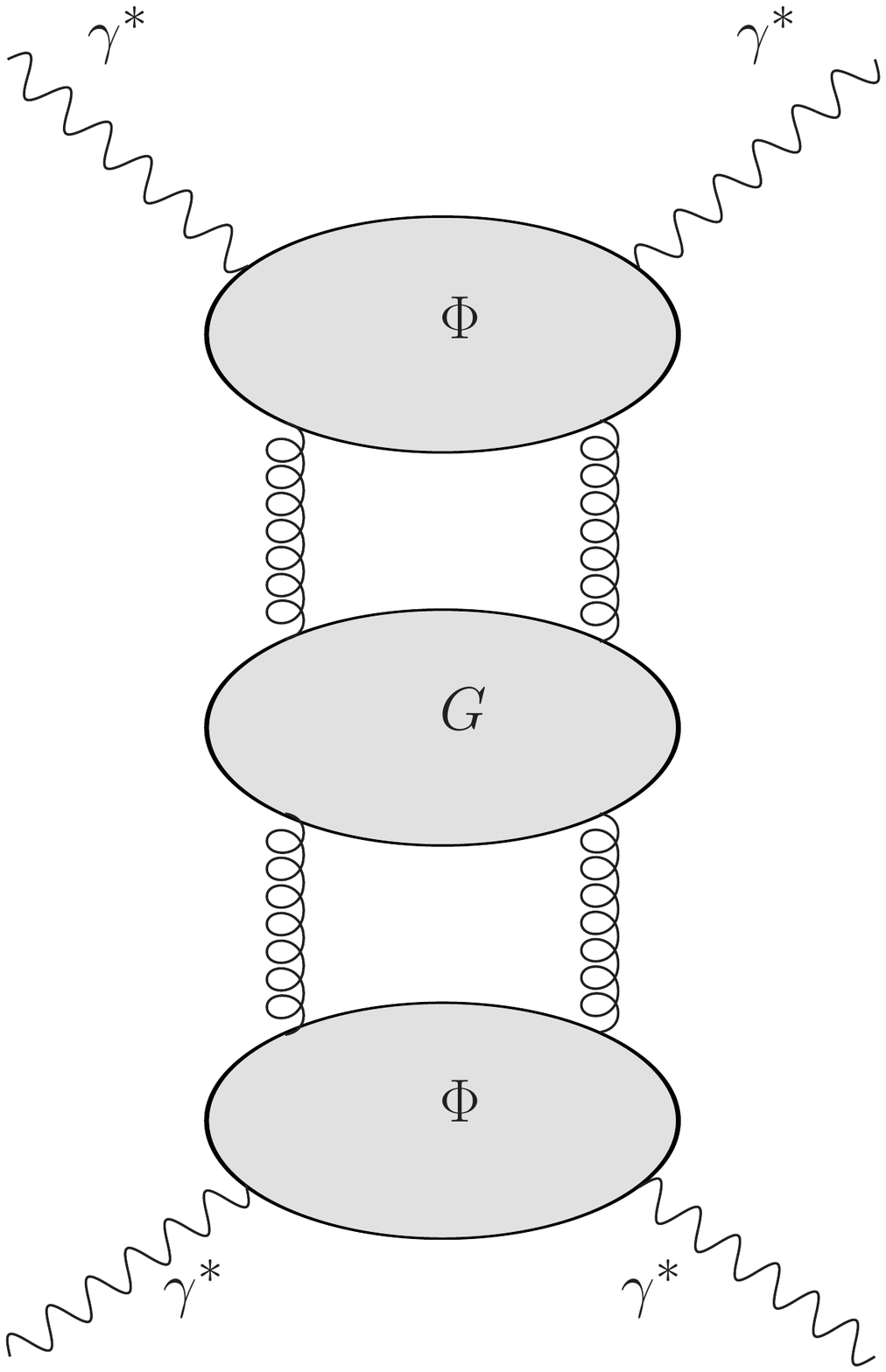}\hspace*{1cm}
				\includegraphics[width=5cm]{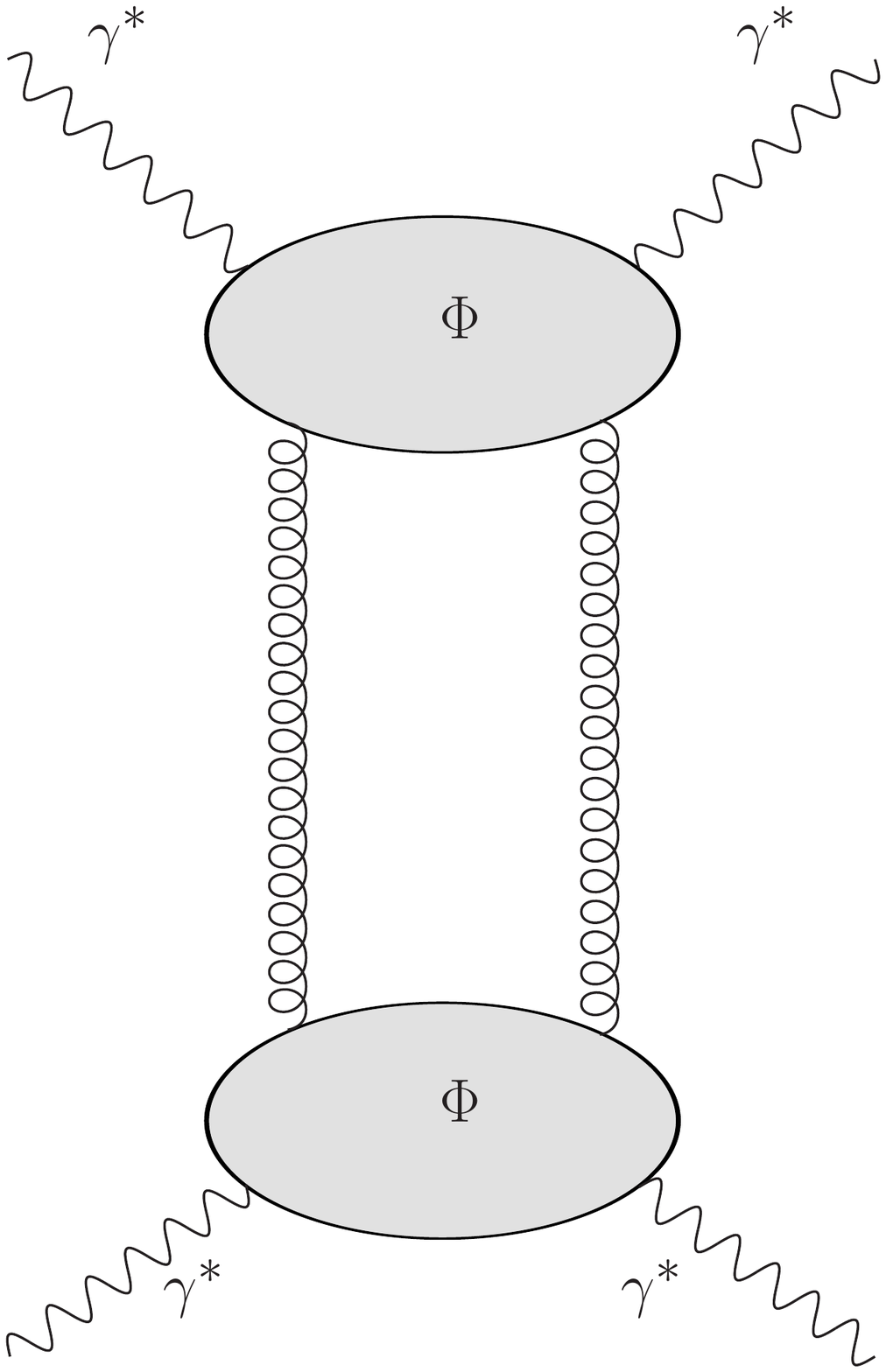}
		\label{fig:gstargstar}
		\caption{Diagram for $\gamma^*\gamma^*$ scattering in the high energy approximation. Left: the diagram with BFKL evolution, right: Born diagram with two gluon exchange.}
	\end{center}
\end{figure}

In the previous section we have discussed the preasymptotic features of the gluon Green's function stemming from the solution to the resummed BFKL evolution within the CCSS scheme. We have identified plateau and even  a dip in the solution as a function of rapidity which depends on the values of the external scales. The basic behavior is such that, as the scales become larger the preasymptotic plateau extends to the larger values of rapidities and it is only at several units of rapidity when the onset of BFKL growth takes place. However, the results of the previous section are rather academic since they do not really tell us anything about the preasymptotic region in the physical processes.  One could expect from the considerations of the previous section that the onset of the BFKL regime will be 
significantly delayed, perhaps  by several units of rapidity. However, the exact position of the minimum and the extent of the plateau will crucially depend on the details of the initial conditions for the solutions, see Fig.~\ref{fig:inputvariation}. In fact, such features  may vary from process to process, given that in the physical reaction the gluon Green's function is convoluted with the impact factor, that allows the coupling of the Pomerons to hadrons on a given process. Even if the impact factor is peaked at some value of the transverse momenta, there will be a significant spread in these momenta.  In order to make more realistic predictions, we shall now investigate a process of the scattering of two virtual photons, producing  light and  heavy quarks. The gluon Green's function needs to be convoluted with the impact factors, which will provide for the suitable initial conditions for the evolution. The predictions will be made for the energies which potentially could be achieved in the future high energy $e^+e^-$ machines, for example like planned FCC-ee \cite{Abada:2019lih,Abada:2019zxq} or CLIC \cite{Battaglia:2004mw,Linssen:2012hp}.  The diagram for the process of the $\gamma^* \gamma^*$ scattering  is indicated in Fig.~\ref{fig:gstargstar}.

This process was calculated in number of works \cite{Brodsky:1996sg,Brodsky:1997sd} using the solution to the BFKL equation in LL order and also  with kinematical constraint \cite{ Kwiecinski:1997ee}.

We shall now recall basic formulae for the two-scale process of scattering of two virtual photons at very high energies. The advantage of such a process is the possibility of the control of the scales in the process and study of the cross section behavior as a function of energy for fixed virtualities or given masses of the produced quarks. In the present analysis we shall use LO impact factors, but with exact kinematics as implemented in \cite{Kwiecinski:1997ee} and \cite{Kwiecinski:1999yx}. This includes partially higher order corrections in the impact factors. Let us denote  by $q_1,q_2$ the four-momenta of the photons. We define $Q_1^2=-q_1^2$ and  $Q_2^2=-q_2^2$ and $W^2=(q_1+q_2)^2$ is the center-of-mass energy squared of the photon-photon system. The cross section for the scattering of two virtual photons due to the exchange of the hard Pomeron has the following form \cite{Kwiecinski:1999yx}

\begin{equation}
\sigma^{\gamma^*\gamma^*}_{ij}(W,Q_1^2,Q^2_2) \; = \; \frac{1}{2\pi} \, \int_{k_0^2}^{k_{\rm max}^2} \frac{dk^2}{k^4} \, \int_{\xi_{\rm min}}^{1/x}d\xi \,  G_i(x\xi,k^2,Q_1^2) \, \Phi_{jq}(k^2,Q_2^2,\xi) \; ,
\label{eq:hardxsection}
\end{equation}
where the limits on the integrals are given by
\begin{equation}
k_{\rm max}^2 = -4m_q^2 + Q_2^2 \bigg(\frac{1}{x}-1\bigg) \; ,
\label{eq:kmax2}
\end{equation}
and
\begin{equation}
\xi_{\rm min} = 1 + \frac{k^2+4m_q^2}{Q_2^2} \; .
\end{equation}
The indices $ij$ denote the two possible polarizations of two virtual photons: $T$ for transverse and $L$ for longitudinal polarization.   The mass of the quark in the quark-box is $m_q$.
The Bjorken $x$ variable is defined as
\begin{equation}
x \;=\;\frac{Q_2^2}{2q_1\cdot q_2} \; .
\label{eq:xbjorkengammagamma}
\end{equation}
Functions $\Phi_{jq}(k^2,Q_2^2,\xi)$ denote the photon-gluon impact factors, i.e. the contributions of the coupling  of the  two-gluon system to the virtual photon through the box and crossed-box diagrams (at LO). Their exact expressions  (see  \cite{Kwiecinski:1997ee, Kwiecinski:1999yx} ) are 
\begin{multline}
\Phi_{Tq}(k^2,Q^2,\xi) \; = \; 2 \alpha_{\rm em} \, \alpha_s(\mu^2)\, e_q^2 \int_0^{\rho_{\rm max}} d\rho \int \frac{d^2 {\mathbf{p}}'}{\pi} \delta \bigg[\xi -\bigg(1+\frac{p^{\prime 2}+m_q^2}{z(1-z)Q^2}+\frac{k^2}{Q^2}\bigg)\bigg] \times  \\
\bigg\{ \bigg[(z^2+(1-z)^2) \bigg(\frac{\mathbf{p}}{D_1}-\frac{\mathbf{p+k}}{D_2}\bigg)^2\bigg]+m_q^2\bigg(\frac{1}{D_1}-\frac{1}{D_2}\bigg)^2  \bigg\}\;,
\label{eq:impact-T}
\end{multline}
for the transverse polarization and
\begin{multline}
\Phi_{Lq}(k^2,Q^2,\xi) \; = \; 8 \alpha_{\rm em} \, \alpha_s(\mu^2)\, e_q^2 Q^2\int_0^{\rho_{\rm max}} d\rho \int \frac{d^2 {\mathbf{p}}'}{\pi} \delta \bigg[\xi -\bigg(1+\frac{p^{\prime 2}+m_q^2}{z(1-z)Q^2}+\frac{k^2}{Q^2}\bigg)\bigg] \times  \\
 \bigg[z^2(1-z)^2 \bigg(\frac{1}{D_1}-\frac{1}{D_2}\bigg)^2\bigg] \;,
 \label{eq:impact-L}
\end{multline}
for the longitudinal polarization. In the above equations the following variables were defined
\begin{eqnarray}
& z & = \frac{1+\rho}{2} \\
& \mathbf{p} & = \mathbf{p}' + (z-1)\mathbf{k}\; ,
\end{eqnarray}
and the denominators
\begin{eqnarray}
& D_1 & = p^2 + z(1-z)Q^2 +m_q^2\; , \\
& D_1 & = (\mathbf{p}+\mathbf{k})^2 + z(1-z)Q^2 +m_q^2 \; ,
\end{eqnarray}
and the maximum value on the $\rho$ integration is 
\begin{equation}
\rho_{\rm max} = \sqrt{1-\frac{4m_q^2}{Q^2(1/x-1)-k^2}} \; .
\end{equation}
The positivity of the argument under the square root leads to the maximum value of $k_{\rm max}^2$ given by \eqref{eq:kmax2}.
The expression \eqref{eq:hardxsection} only contains the contribution due to the exchange of the hard Pomeron, i.e. the perturbative part. In principle the cross section can contain some non-perturbative components which should be modeled. 
For example, the integrals in \eqref{eq:hardxsection}  have a lower cutoff $k_0^2$, which can be interpreted as an non-perturbative cutoff. For the phenomenology, these contributions need to be modeled, and they can be numerically important for low energies $W<100 \; \rm GeV$ as discussed in  \cite{Kwiecinski:1999yx}.
 We shall investigate the sensitivity of the results to the cutoff parameter in detail later on in this section.

The function $G_i(k^2,Q^2,x\xi)$ in \eqref{eq:hardxsection} is the solution to the BFKL equation with the initial condition given by
\begin{multline}
G_T^0(k^2,Q^2,x) = 2 \alpha_{\rm em} \alpha_s(\mu^2)\sum_q e_q^2 \int_x^1 dz  \int_0^1 d\lambda \left\{  \frac{[\lambda^2+(1-\lambda)^2][z^2+(1-z)^2]k^2}{\lambda(1-\lambda)k^2+z(1-z)Q^2+m_q^2} \right.   \\ \left.
+2m_q^2 \left[\frac{1}{z(1-z)Q^2+m_q^2}-\frac{1}{\lambda(1-\lambda)k^2+z(1-z)Q^2+m_q^2}\right] \right\} \; ,
\label{eq:initial-T}
\end{multline}
for transverse polarization and
\begin{multline}
G_L^0(k^2,Q^2,x) = 16 \alpha_{\rm em} \alpha_s(\mu^2) Q^2 k^2 \sum_q e_q^2 \int_x^1 dz  \\  \int_0^1 d\lambda \left\{  
\frac{\lambda(1-\lambda)z^2(1-z)^2}{[\lambda(1-\lambda)k^2+z(1-z)Q^2+m_q^2][z(1-z)Q^2+m_q^2]} 
 \right\} \; ,
\label{eq:initial-L}
\end{multline}
for the longitudinal polarization.
The above expressions can be obtained from functions \eqref{eq:impact-T} and \eqref{eq:impact-L} by integration  and they again correspond of the coupling of two gluons to the quark box.

We have implemented the CCSS resummed solution (with contributions up to NLLx BFKL and NLO DGLAP) as presented in the previous section to calculate the $\gamma^* \gamma^*$ cross section using formulae presented in  Eqs.~\eqref{eq:hardxsection}-\eqref{eq:initial-L}.

\begin{figure}
	\begin{center}
		\includegraphics[width=0.49\textwidth]{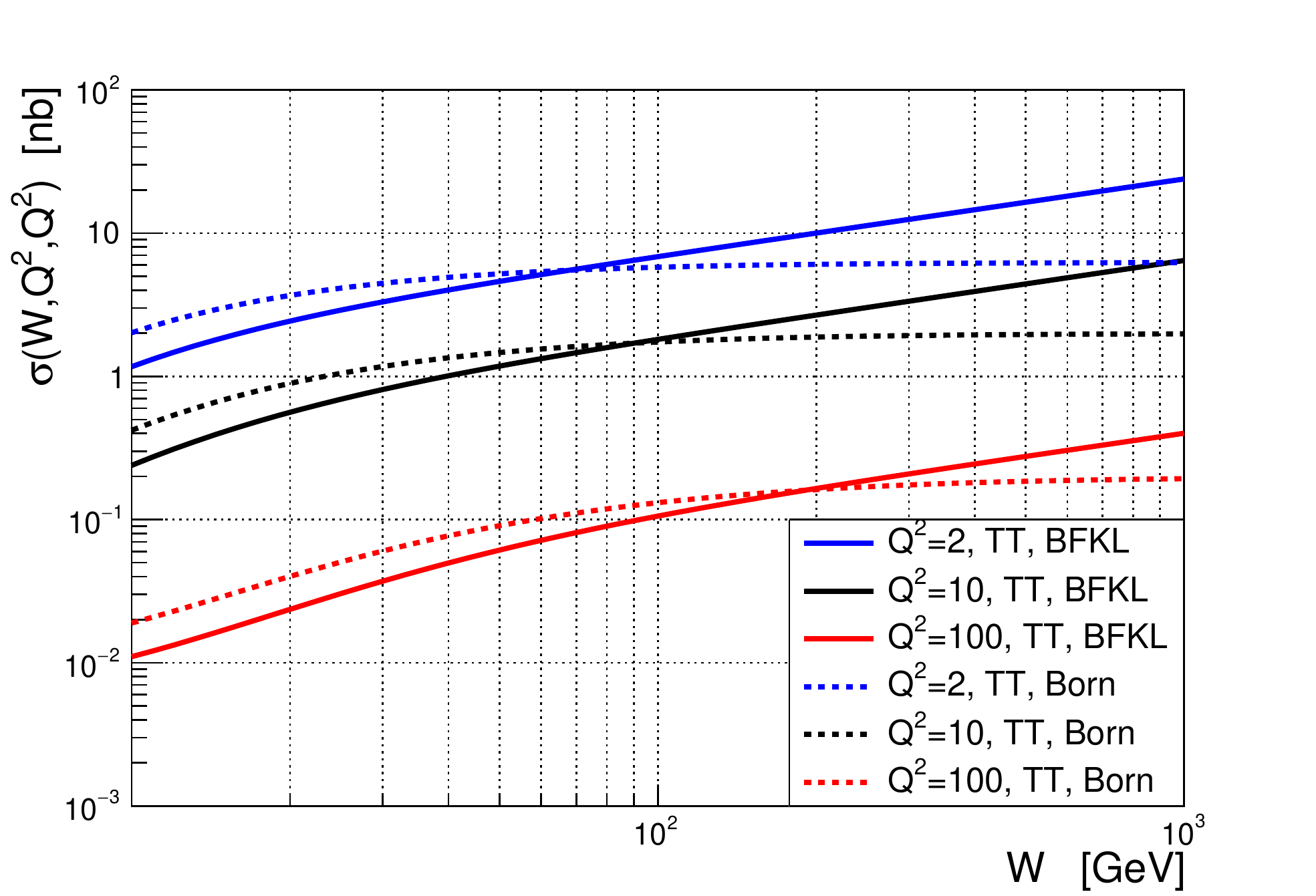}
		\includegraphics[width=0.49\textwidth]{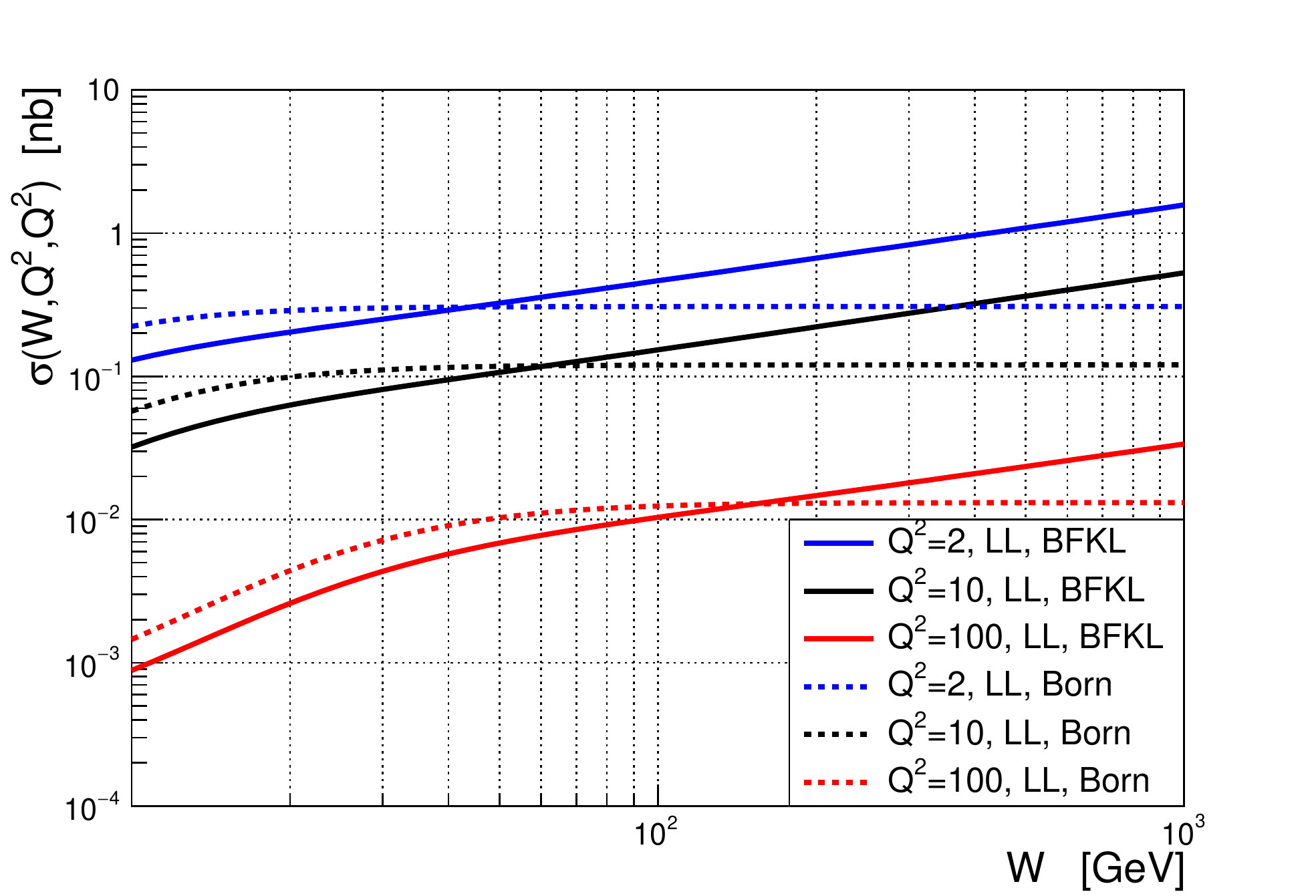}
	\end{center}
	\caption{The cross section $\sigma^{\gamma^*\gamma^*}(W,Q^2,Q^2)$ for virtual $\gamma^*\gamma^*$ scattering as a function of energy $W$ for three different virtualities of the photons $Q^2=Q_1^2=Q_2^2=2,10,100 \; \rm GeV^2$ (blue, black, red lines).
		Solid lines: BFKL calculation with resummation; dashed lines:  Born calculation. Left plot:  polarizations of both photons are transverse. Right plot: polarization of both photons are longitudinal.}
	\label{fig:sigmaW}
\end{figure}

In Fig.~\ref{fig:sigmaW} we show the cross section as a function of energy $W$ and for the selected virtualities $Q_1^2,Q_2^2$ of both photons. In all calculations we assumed the virtualities of the photons to be equal, i.e. $Q_1^2=Q_2^2=Q^2$. We consider  two cases, where the two photons have the same polarizations: transverse-transverse or longitudinal-longitudinal. There is also longitudinal-transverse case, whose magnitude is between the two cases.  The calculation is done for three massless flavors and a massive  charm quark with  a mass of $m_c=1.27\, \rm GeV$. The solid lines indicate the cross section based on the resummed BFKL
 evolution. The dashed lines correspond to the Born calculation, i.e. two gluon exchange between impact factors, which is obtained by  using \eqref{eq:initial-T} and \eqref{eq:initial-L} in \eqref{eq:hardxsection} for $G_L,G_T$.  
The Born calculation therefore tends to a constant value at highest values of the energy $W$. The initial  growth of the Born cross section at low and intermediate energies, $ W \lesssim 100 \; {\rm GeV}$,  is due to the opening of the  phase space for the impact factors.
We see a clear growth of the BFKL calculation with $W$, which leads to a much higher cross section than the Born exchange for highest energies, $W > 100 \, {\rm GeV}$. However,
the BFKL calculation  is {\em lower} than the Born calculation for low energies, of the order of $W \simeq 10-100 \; {\rm GeV}$. The point where the cross section with the BFKL resummation becomes larger than the Born calculation depends strongly on the virtualities of the photons, i.e. for higher virtualities it moves to a higher energies.

\begin{figure}
	\begin{center}
			\includegraphics[width=0.48\textwidth]{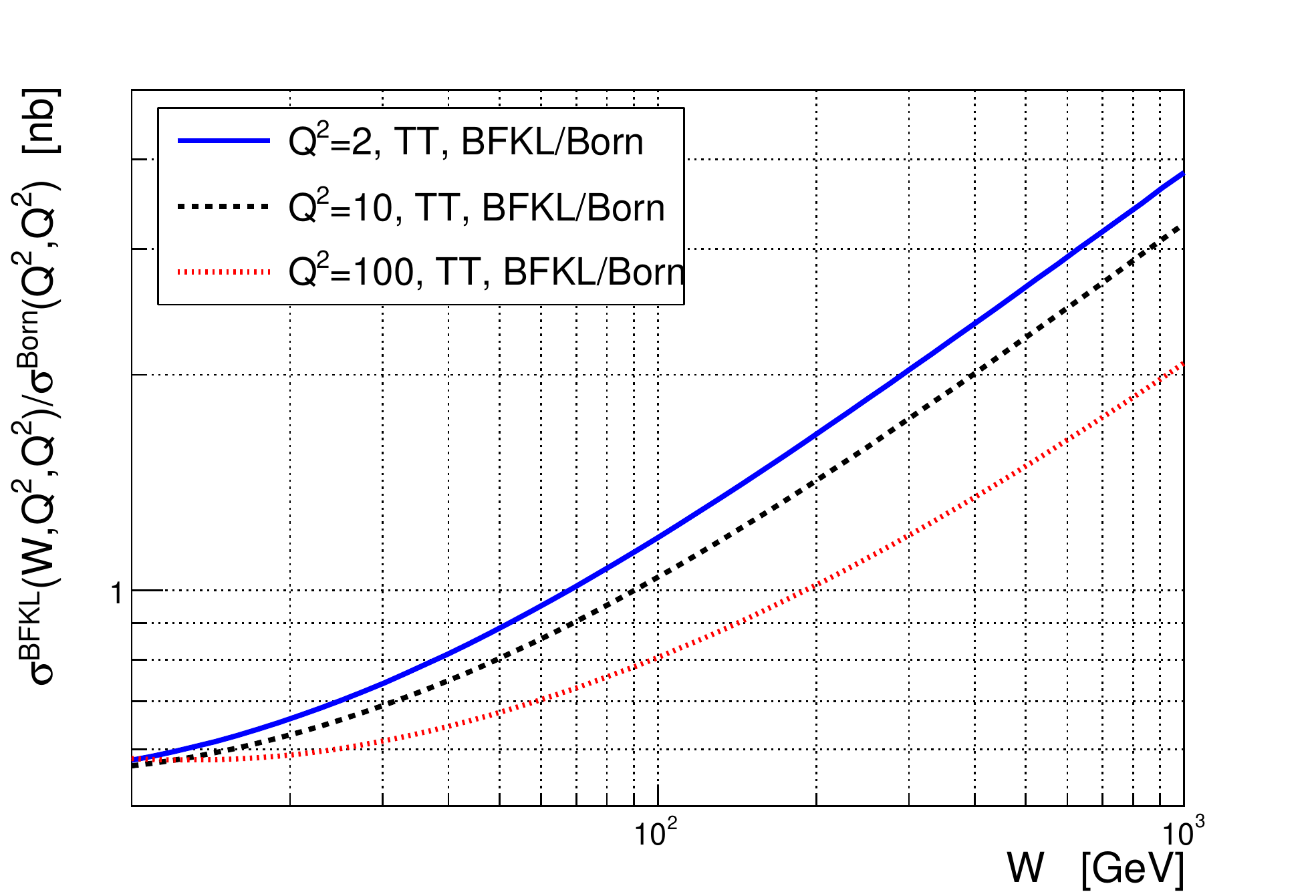}
		\includegraphics[width=0.48\textwidth]{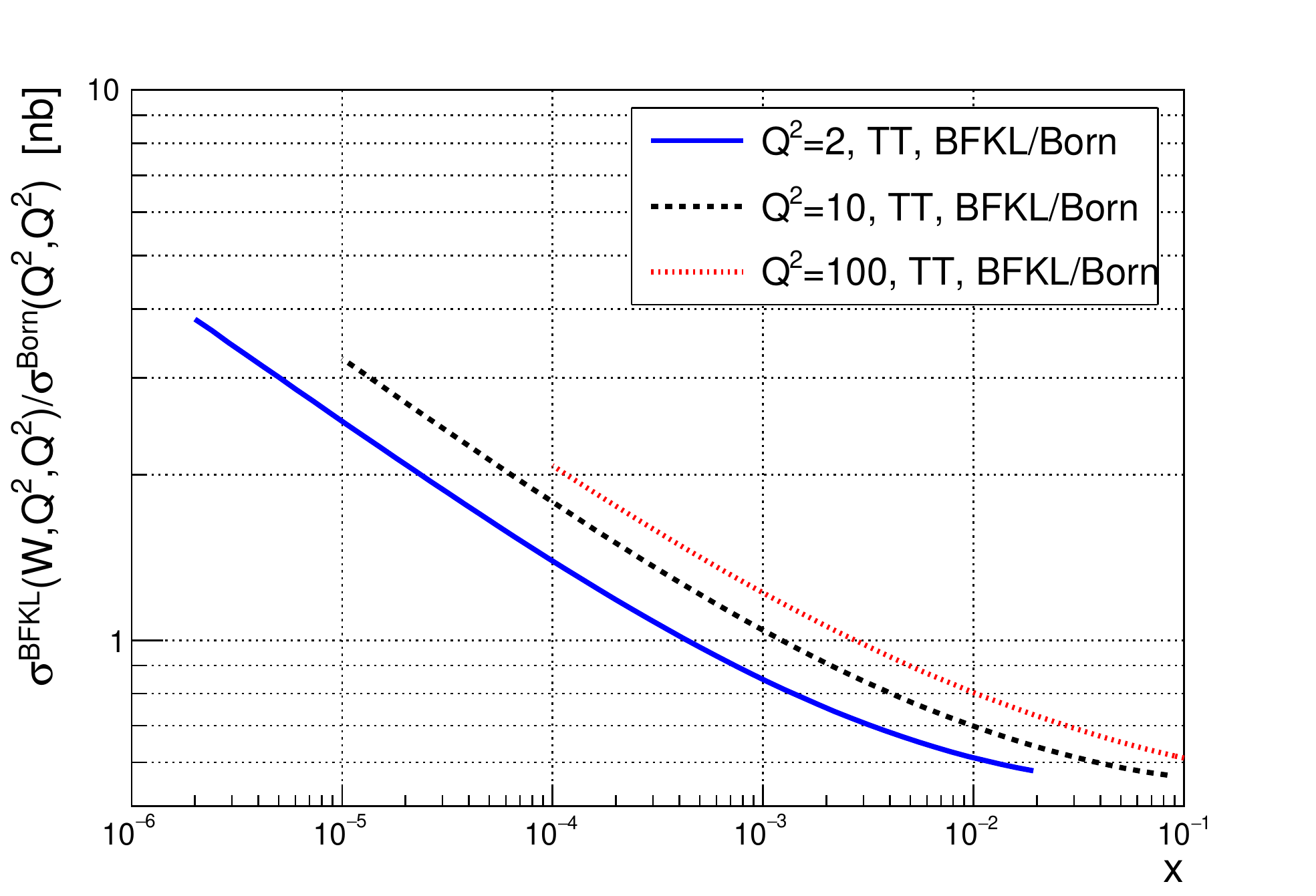}
	\end{center}
	\caption{The cross section ratio of resummed BFKL to Born for virtual $\gamma^*\gamma^*$ scattering as a function of energy $W$(left) and $x$(right) for different virtualities of the photons $Q^2=Q_1^2=Q_2^2=2,10,100 \; \rm GeV^2$.
		Polarizations of both photons were  taken to be transverse.
	}
	\label{fig:ratioBFKLtoBornW-T}
\end{figure}

\begin{figure}
	\begin{center}
		\includegraphics[width=0.48\textwidth]{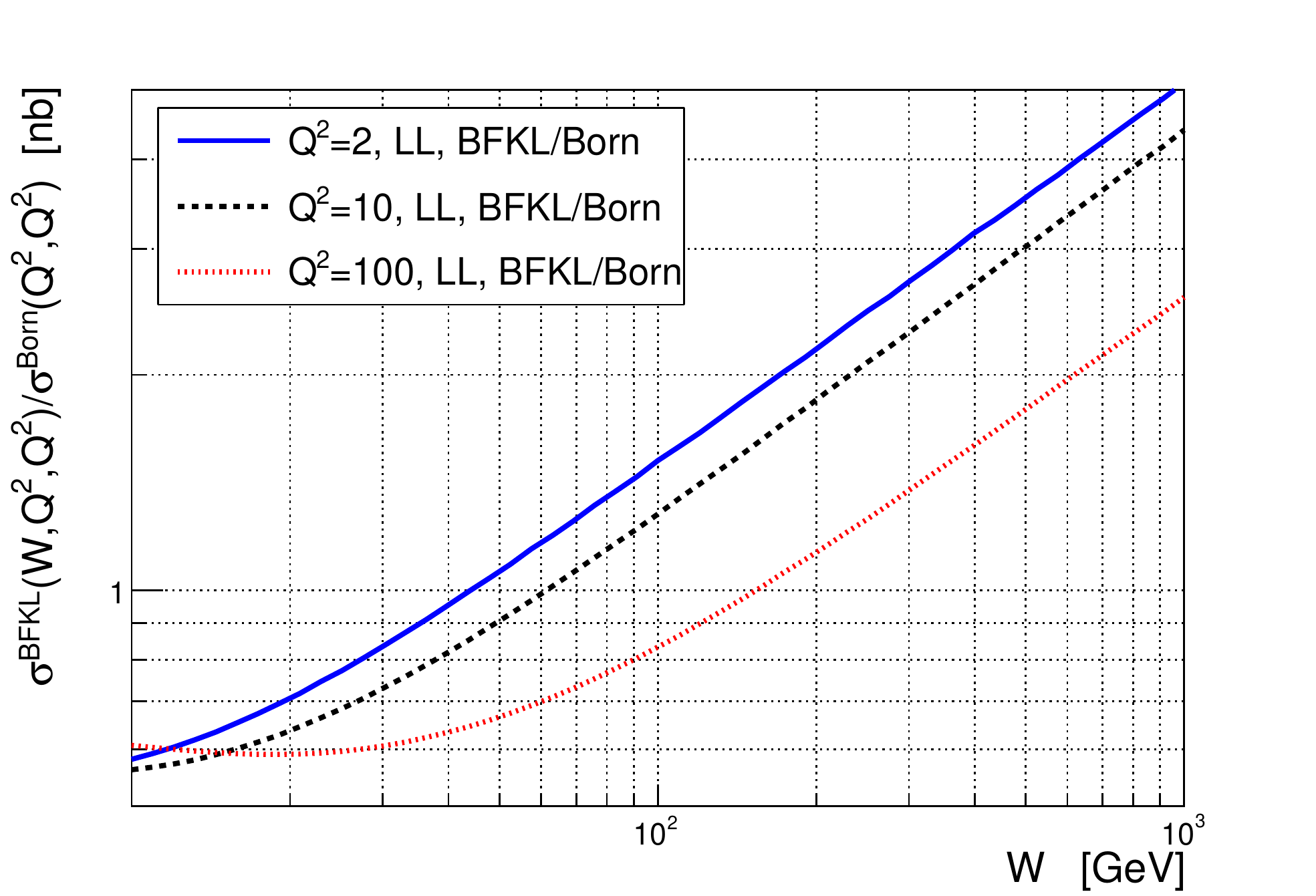}
		\includegraphics[width=0.48\textwidth]{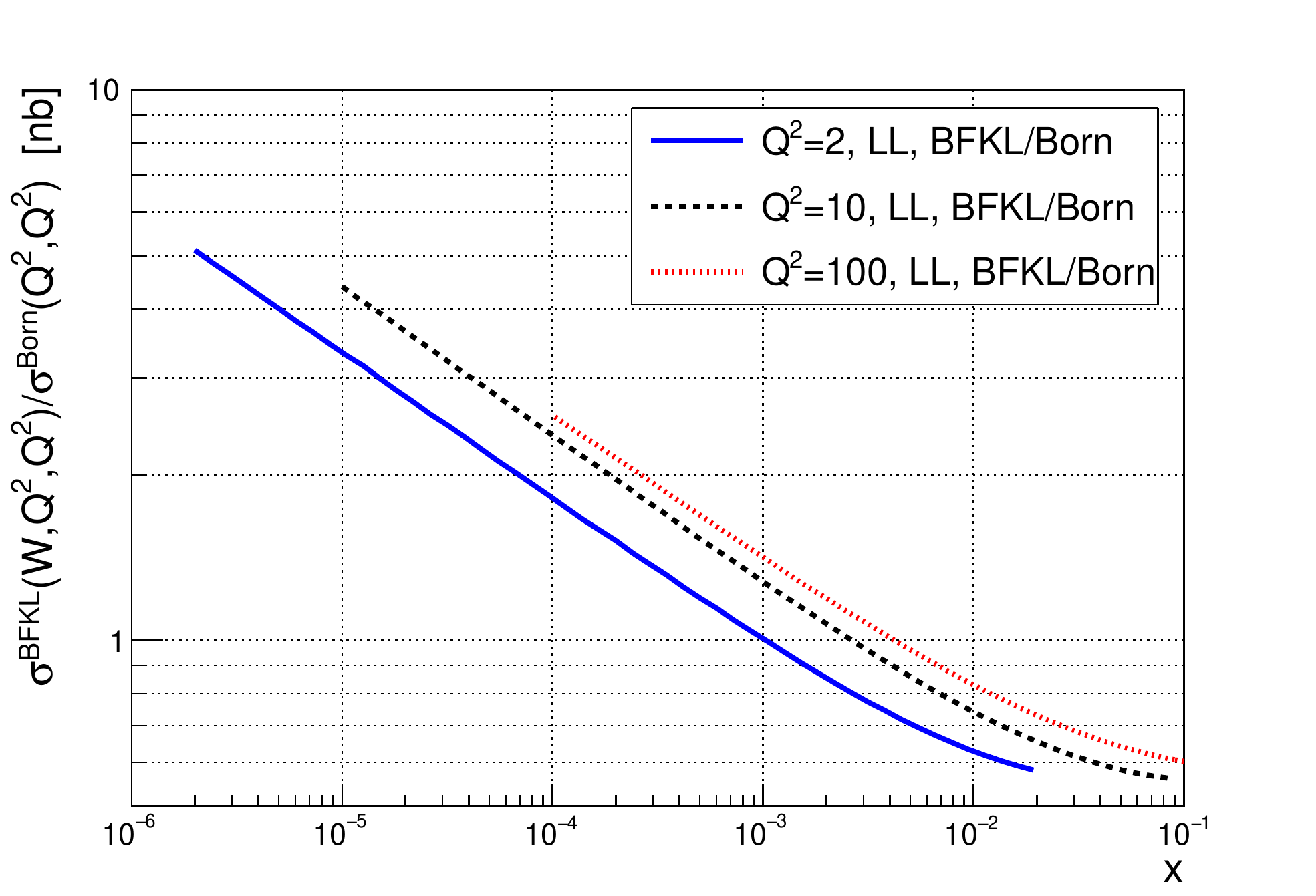}
	\end{center}
	\caption{The cross section ratio of resummed BFKL to Born for virtual $\gamma^*\gamma^*$ scattering as a function of energy $W$ (left) and $x$(right) for different virtualities of the photons $Q^2=Q_1^2=Q_2^2=2,10,100 \; \rm GeV^2$.
		Polarizations of both photons were  taken to be longitudinal.
	}
	\label{fig:ratioBFKLtoBornW-L}
\end{figure}

\begin{figure}
	\begin{center}
		\includegraphics[width=0.52\textwidth]{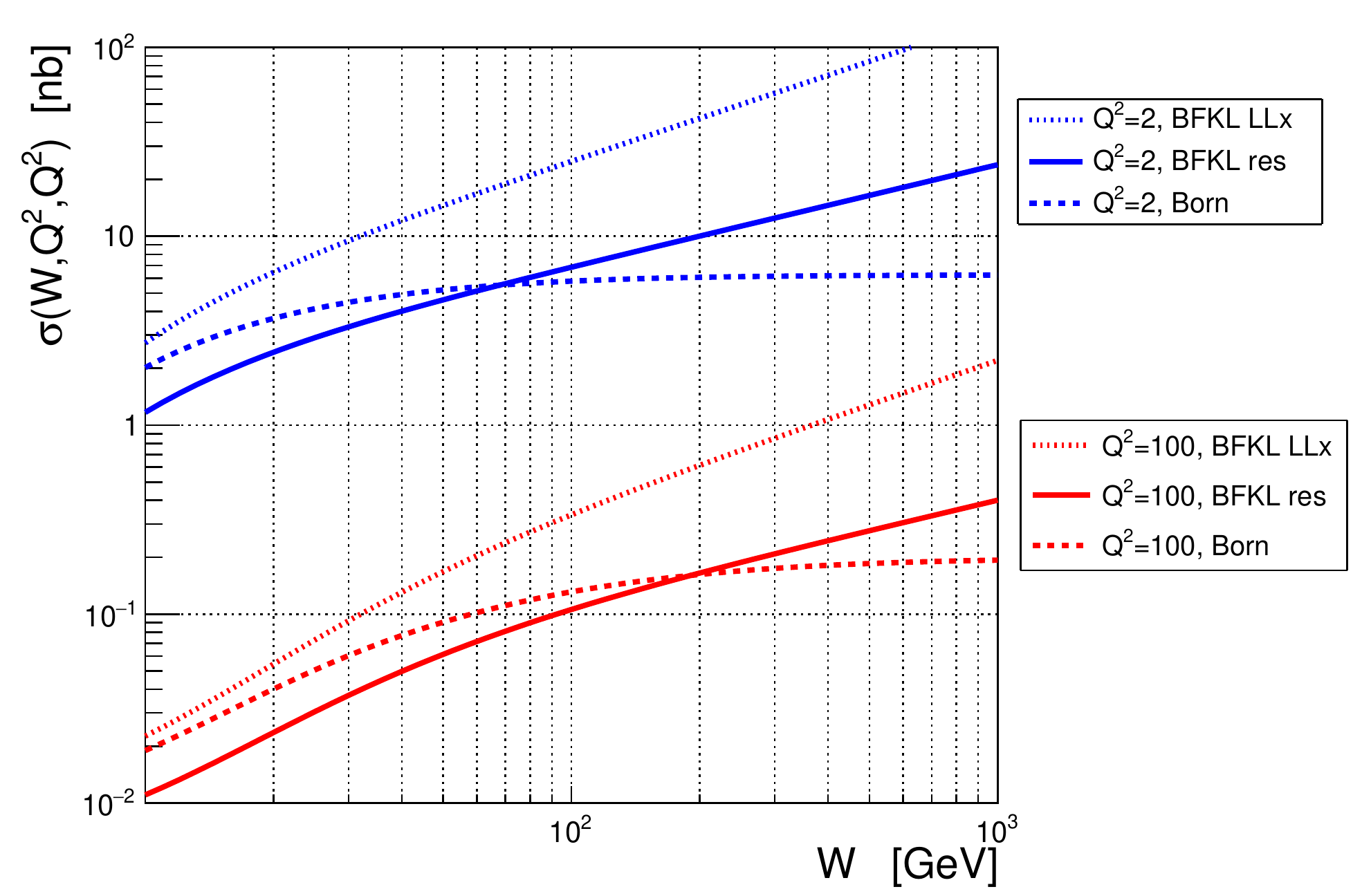}
		\includegraphics[width=0.47\textwidth]{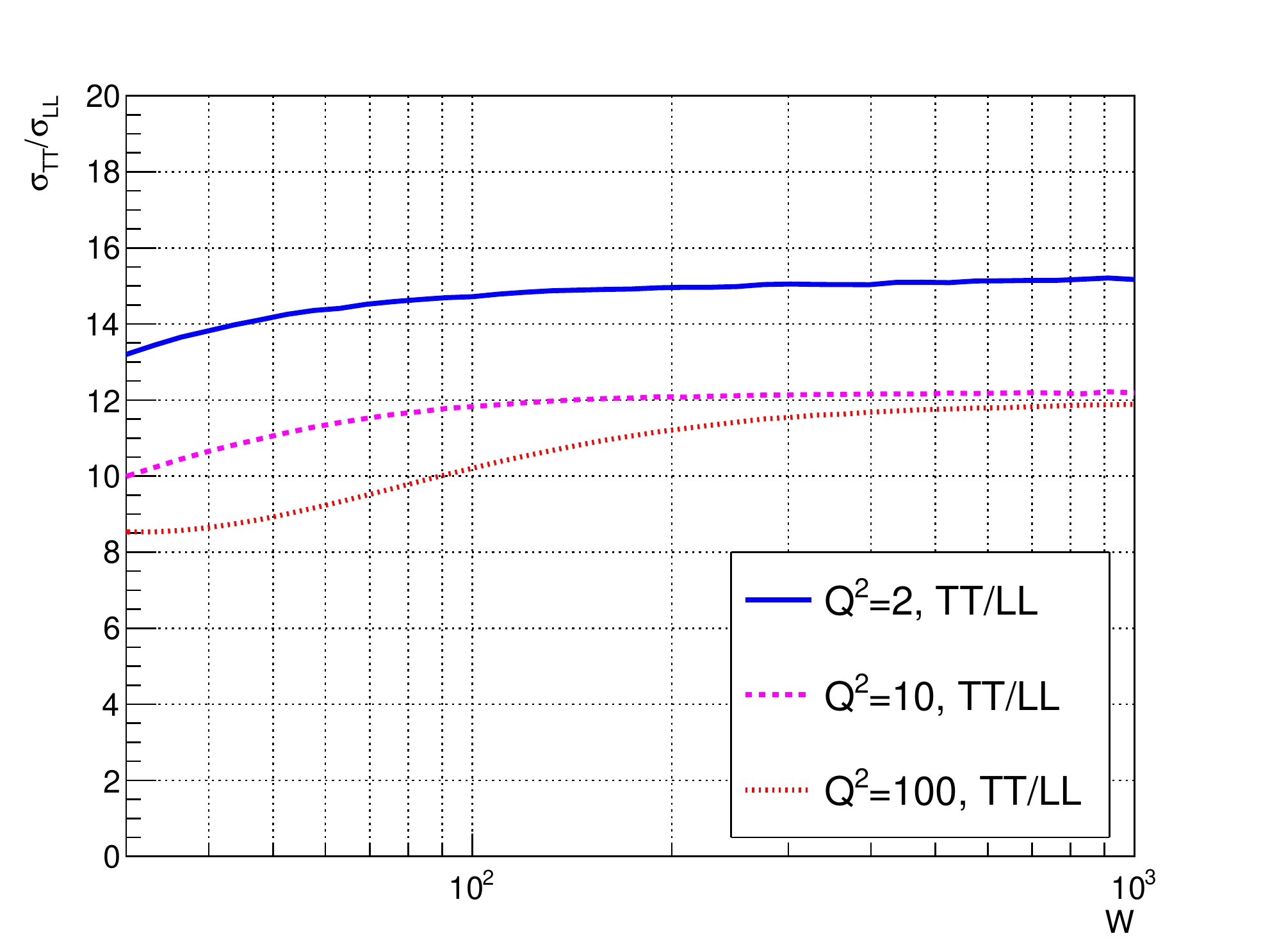}
	\end{center}
	\caption{Left: the transverse-transverse cross section as a function of energy $W$ for two values of $Q^2=2$ and $100 \; \rm GeV^2$. The BFKL resummed calculation is compared to the LLx BFKL calculation with the running coupling, and the Born case. Right: the cross section ratio of transverse to longitudinal polarization  as a function of energy $W$ for different virtualities of the photons $Q^2=Q_1^2=Q_2^2=2,10,100 \; \rm GeV^2$.
	}
	\label{fig:ratioTTLL}
\end{figure}

This effect is better illustrated in  Figs.~\ref{fig:ratioBFKLtoBornW-T} for the transverse-transverse case and Fig.~\ref{fig:ratioBFKLtoBornW-L} for the longitudinal-longitudinal case where the ratio of the resummed BFKL to Born calculation is shown as a function of energy $W$ on the left plots, and as a function of $x$ variable on the right plots.
We see that the BFKL based computation is in fact much lower at low energies than the Born calculation.
 One could argue that perhaps the BFKL calculation should not be completely trusted at very large values of $x \sim 0.1$ and above, nevertheless the effect is prominent even at values of $x$ which are lower  than $0.01$.
 This  effect, as well as a delay in the onset of the asymptotic small $x$ growth, was observed also in \cite{Kwiecinski:1999yx} which used kinematical constraint in the LL BFKL with running coupling, and thus partially accounting for the resummation effects.  In the present calculation, which uses full resummation,
 the effect is more pronounced. Most likely this is due to the additional non-leading effects incorporated into resummation, like the non-singular parts of the DGLAP splitting function which tend to bring the solution even lower. We see that the preasymptotic region extends to quite large energies, of the order of $50-200 \, \rm GeV$, or small $x \sim 5 \div 1 \times 10^{-3}$, depending on the virtuality. We have also verified that the calculation based on the LL BFKL equation (with just running coupling included) is always larger than the Born calculation, even for low energies.  This is illustrated in Fig.~\ref{fig:ratioTTLL} (left plot). We thus conclude that the strong suppression of the calculation based on the resummed evolution  with respect to the Born calculation is indeed due to the non-leading effects and resummation. It is thus  connected to the strong preasymptotic effects of the  gluon Green's function analyzed in the previous section.

We observe,  that even though the resummed  BFKL cross section is lower than the Born cross section for lower energies, the ratio still grows with energy, starting from the low energies.
Only in the case of $Q^2=100 \, \rm GeV^2$, particularly for the longitudinal-longitudinal case, a visible delay in the onset of the growth is present, similar to the plateau for the gluon Green's function analyzed in the previous section.  This shows that when the gluon Green's function is convoluted with the impact factor, the dependence on the energy (particularly at lowest energies) is strongly modified with respect to the `bare' gluon Green's function.

We also note that for the case of longitudinal-longitudinal polarization the BFKL solutions dominates the Born calculation, i.e. starting with lower energies, than in the transverse-transverse case (compare Figs.~\ref{fig:ratioBFKLtoBornW-T} and \ref{fig:ratioBFKLtoBornW-L}).

This is related to the difference in the energy behavior of the two polarization cases. In Fig.~\ref{fig:ratioTTLL},  we show the ratio of the transverse-transverse contribution to the longitudinal-longitudinal case. The transverse-transverse contribution dominates the cross section, and the ratio is rather flat for the energies considered.  We checked that, for the higher values of $Q^2$ the ratio  starts to slightly decrease for higher energies $W$. Overall, the LL contribution is about $10\%$ of the TT contribution for the range of parameters explored. 

We also observe, from Fig.~\ref{fig:sigmaW} and  Fig.~\ref{fig:ratioBFKLtoBornW-T}  that the exponent  governing the energy growth for the transverse-transverse case is almost identical for all the values of photon virtualities.  From the analysis of the previous section, and in general expectation from the solutions to the BFKL equation one would conclude that the exponent should depend strongly on the value of $Q^2$, since the Pomeron intercept is a function of $\alpha_s$ which should be dominated by the typical values of transverse momenta of the order of virtuality in the impact factor. In such scenario, one would expect that the rate of growth at $Q^2=2 \, \rm GeV^2$ would be  faster than at $Q^2=100 \, \rm GeV^2$, i.e. the ratio of the intercepts would be approximately $\sim 1.5-1.7$, see Fig.~1 in \cite{Ciafaloni:2003rd}. For the transverse-transverse calculation, this does not seem to be the case. The situation is slightly different for the longitudinal-longitudinal calculation. In Fig.~\ref{fig:ratioBFKLtoBornW-L}, particularly in the right plot which shows the $x$ dependence,  the growth with $x$ for $Q^2=100 \, \rm GeV^2$ is a bit slower than for $Q^2 =2$ or $Q^2=10 \, \rm GeV^2$. The latter two, exhibit though similar $x$ dependence.  Overall, there does not seem to be a lot of dependence of the rate of growth with $W$  on the scales of $Q^2$ for the impact factors.

To better understand the behavior of the cross section, and in particular which transverse momenta contribute to the integral, we analyzed the distribution in the transverse momenta in the expression for the cross section Eq.\eqref{eq:hardxsection}.  That is we define the function which is the integrand in Eq.\eqref{eq:hardxsection}
$\bar{p}(Q_1^2,Q_2^2,W,k^2) $  as
\begin{equation}
\sigma_{ij}^{\gamma^*\gamma^*}(W,Q_1^2,Q_2^2) = \int_{k_0^2}^{k_{\rm max}^2} \frac{dk^2}{k^2} {\overline p}_{ij}(W,Q_1^2,Q_2^2,k^2) \; ,
\label{eq:xsecintegrand}
\end{equation}
and plot the normalized distribution
\begin{equation}
p_{ij}(W,Q_1^2,Q_2^2,k^2)= \frac{{\overline p}_{ij}(W,Q_1^2,Q_2^2,k^2)}{\sigma^{\gamma^*\gamma^*}(W,Q_1^2,Q_2^2)} \; ,
\label{eq:xsecintegrandnorm}
\end{equation}
for which the areas under the curves are equal to unity.
\begin{figure}
	\includegraphics[width=0.49\textwidth]{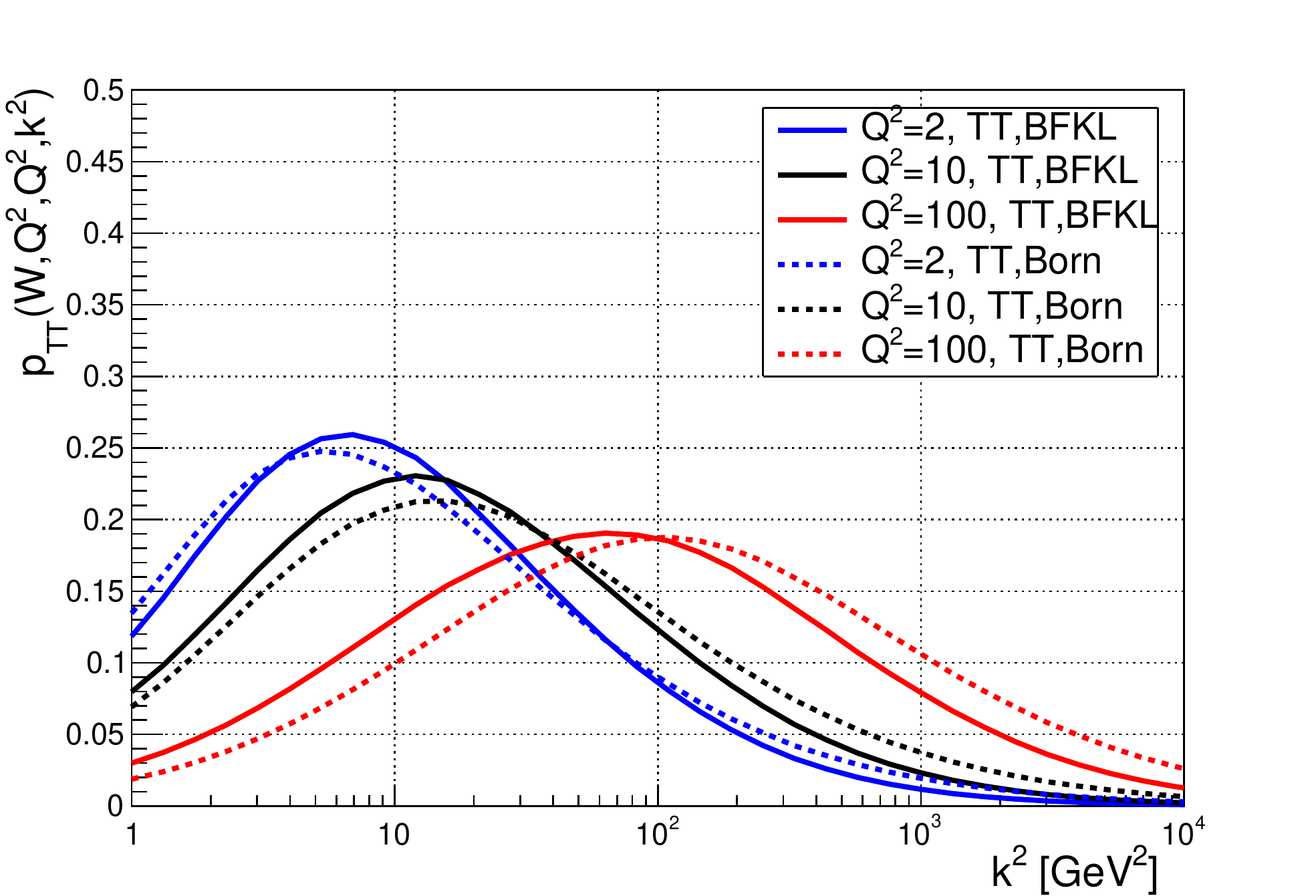}
		\includegraphics[width=0.49\textwidth]{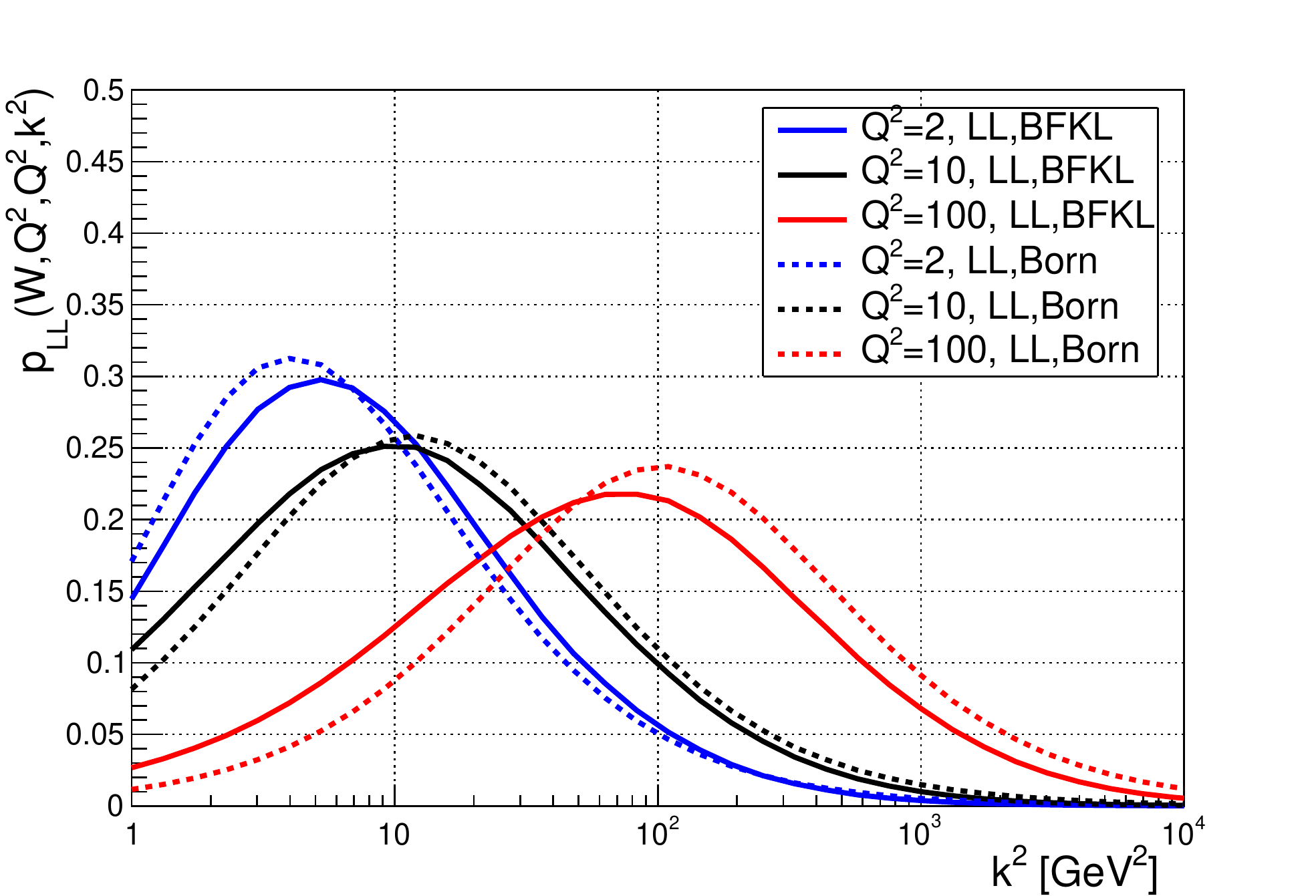}
	\caption{Normalized integrand $p(Q_1^2,Q_2^2,W,k_{T}^2)$ see Eq.\eqref{eq:xsecintegrand} and Eq.\eqref{eq:xsecintegrandnorm} for selected values of external scales $Q^2=Q_1^2=Q_2^2=2,10,100 \, \rm GeV^2$ (blue,black,red curves correspondingly). Dashed lines -- Born calculation, solid lines--BFKL calculation. Left plot: transverse-transverse polarization; right plot: longitudinal-longitudinal polarization. The coupling in the impact factors and in the BFKL evolution has been kept fixed, $\alpha_s=0.2$. }
	\label{fig:pintegrand_fxd}
\end{figure}

\begin{figure}
	\includegraphics[width=0.49\textwidth]{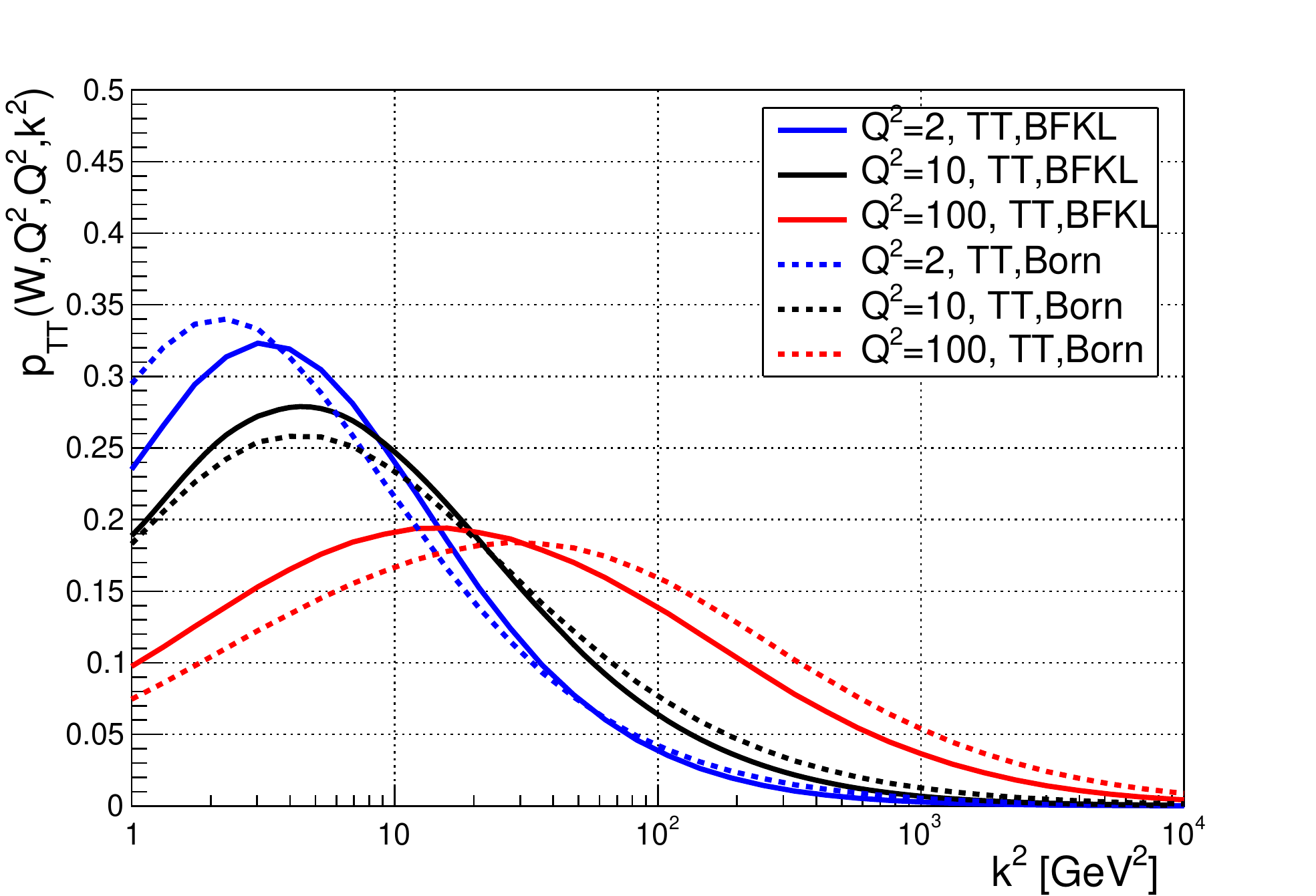}
		\includegraphics[width=0.49\textwidth]{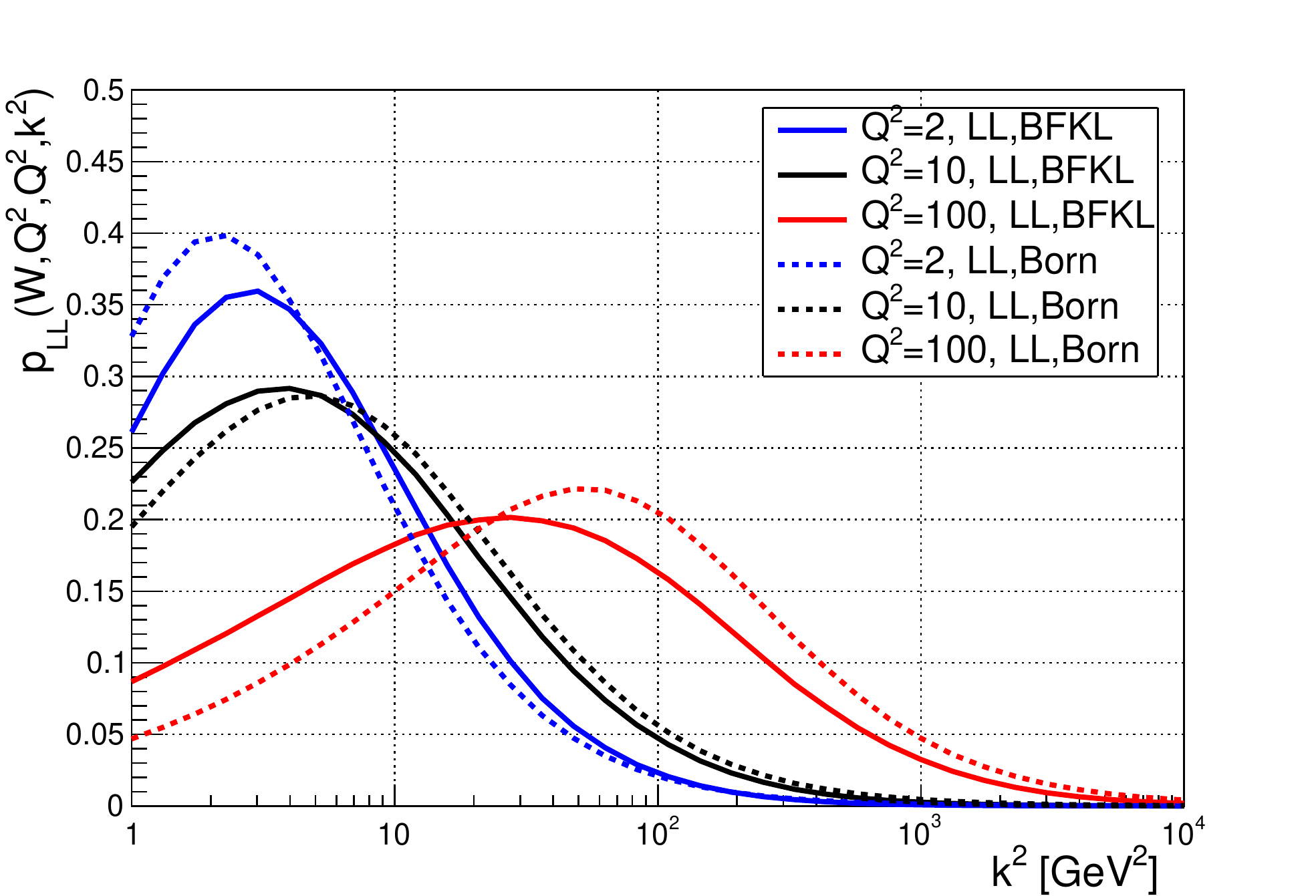}
	\caption{Normalized integrand $p(Q_1^2,Q_2^2,W,k_{T}^2)$ see Eq.\eqref{eq:xsecintegrand} and Eq.\eqref{eq:xsecintegrandnorm} for selected values of external scales $Q^2=Q_1^2=Q_2^2=2,10,100 \, \rm GeV^2$ (blue,black,red curves correspondingly). Dashed lines -- Born calculation, solid lines--BFKL calculation. Left plot: transverse polarization; right plot: longitudinal polarization. The coupling in the impact factors and in the BFKL evolution is running. The scale in the impact factor coupling is $k^2+m_q^2$. }
	\label{fig:pintegrand_run}
\end{figure}

In Figs.~\ref{fig:pintegrand_fxd} and \ref{fig:pintegrand_run} we show these distributions for different sets of $Q^2$ values. In Fig.~\ref{fig:pintegrand_fxd}
we show the calculation with the fixed value of the strong coupling, equal $\bar{\alpha}_s=0.2$, both in impact factors and in the BFKL evolution, indicated by the solid lines. For comparison we show also the calculation for the Born case, i.e. without any evolution, just a two gluon exchange.  The right plot shows the case for the longitudinal-longitudinal polarization, and the left plot is the case for the transverse - transverse polarization. First, we note that the distribution in the longitudinal-longitudinal case is narrower than in the transverse-transverse case. This is  a  well known effect,  mainly due to the presence of the aligned jet configurations in the impact factors, i.e. when $z=0$  or $z=1$ in the case of the transverse polarization (see Eqs. \eqref{eq:initial-T}, \eqref{eq:initial-L}). Second, the peak of the distribution is around the value of $Q^2$. The exception seems to be in the case of very low $Q^2=2 \; \rm GeV^2$, where the distribution is shifted to the slightly higher values of $k^2$.  We note however, that the integrals over $p^2$ in \eqref{eq:impact-T}, \eqref{eq:impact-T} are cut off at $p_0=k_0=1 \; \rm GeV$.

The calculations are strongly modified by the presence of the running coupling.  We note that the running coupling is present both in the BFKL evolution as well as in the impact factors. Following \cite{Kwiecinski:1999yx} we considered two scenarios of the running coupling in the impact factors, with the choice of scale $(k^2+m_q^2)$ and a lower scale $(\frac{k^2+m_q^2}{4})$. The difference between the two can be treated as the scale variation in the impact factor. As evident from Fig.\ref{fig:pintegrand_run} the distribution is strongly shifted to the  lower values of $k^2$ even for large values of $Q^2$. The effect is more prominent in the BFKL case where the diffusion in $k^2$ provides additional shift of the distribution towards infrared regime. As a result, in the case of $Q^2=100 \, \rm {GeV^2}$, the peak of the distribution in $k^2$ is moved down to about $10 \, {\rm GeV^2}$.  
\begin{figure}
	\includegraphics[width=0.49\textwidth]{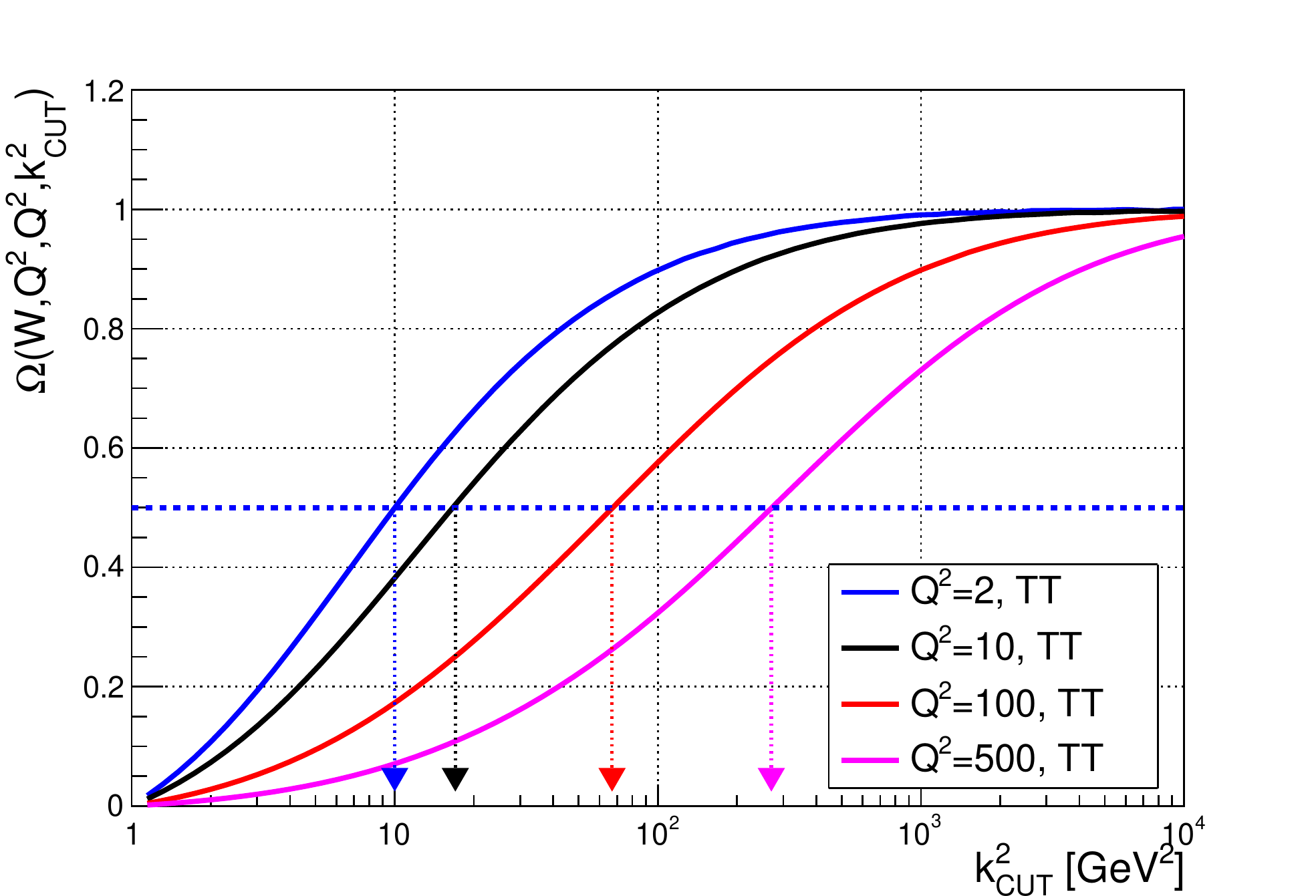}
	\includegraphics[width=0.49\textwidth]{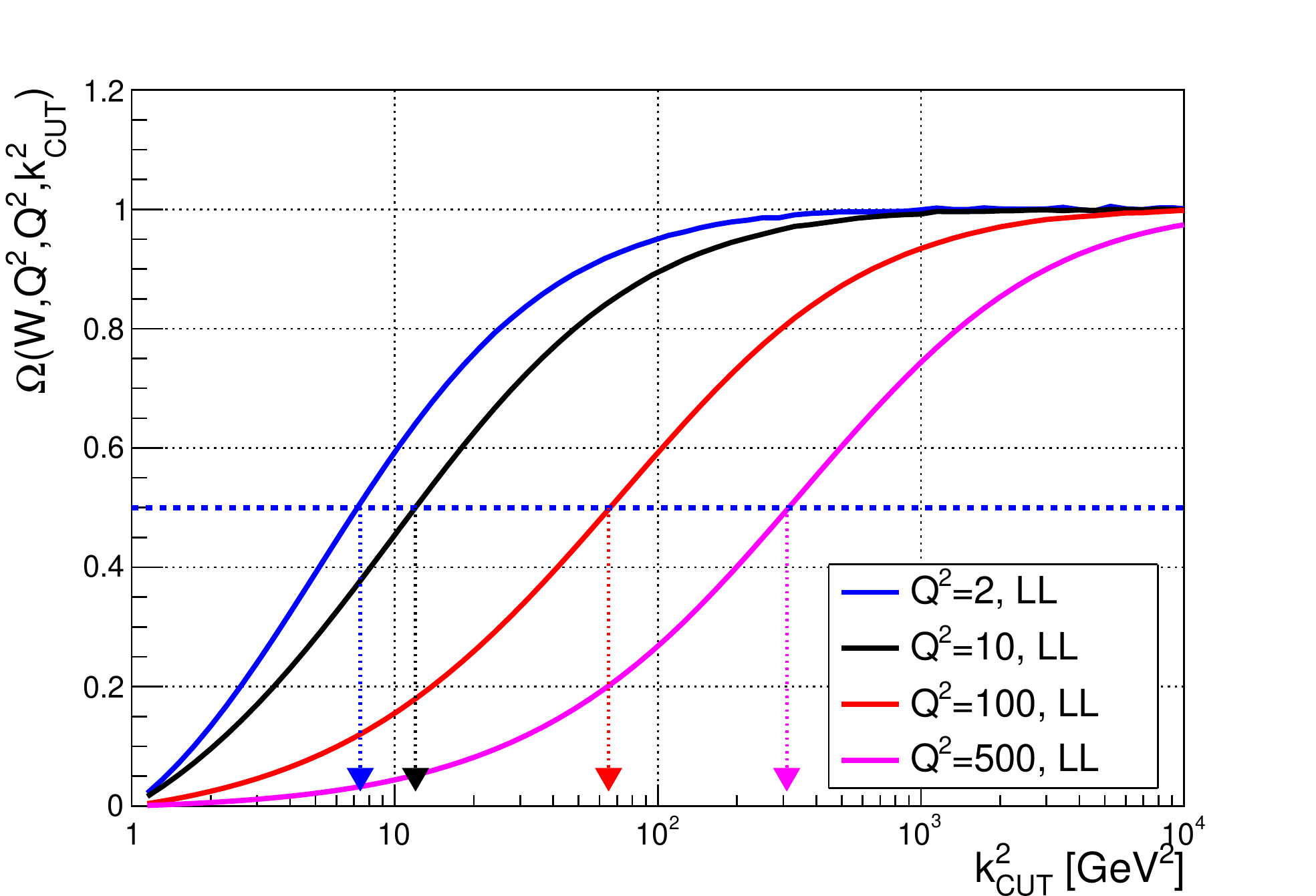}
	\caption{The ratio of the cross section for $\gamma^* \gamma^*$ scattering integrated up to a cut off $k_{\scaleto{\rm CUT}{4pt}}$ over the cross section integrated over the entire range of transverse momenta as a function of $k_{\scaleto{\rm CUT}{4pt}}$. The arrow indicate the positions of the values of $k_{\scaleto{\rm CUT}{4pt}}$  --median values -- for which the ratio is equal to $1/2$. Different curves indicate the different values  of $Q^2$: $2,10,100,500 \; {\rm GeV}^2$ -- blue, black, red and magenta lines correspondingly. The calculation was done for fixed value of the coupling constant, both in impact factors and in the BFKL evolution  $\alpha_s = 0.2$ and for the case of longitudinal-longitudinal (right plot) and transverse-transverse (left plots) polarization of both photons. Only light quark contribution is included in the calculation.}
	\label{fig:ktmedian_fxd}
\end{figure}
\begin{figure}
	\includegraphics[width=0.49\textwidth]{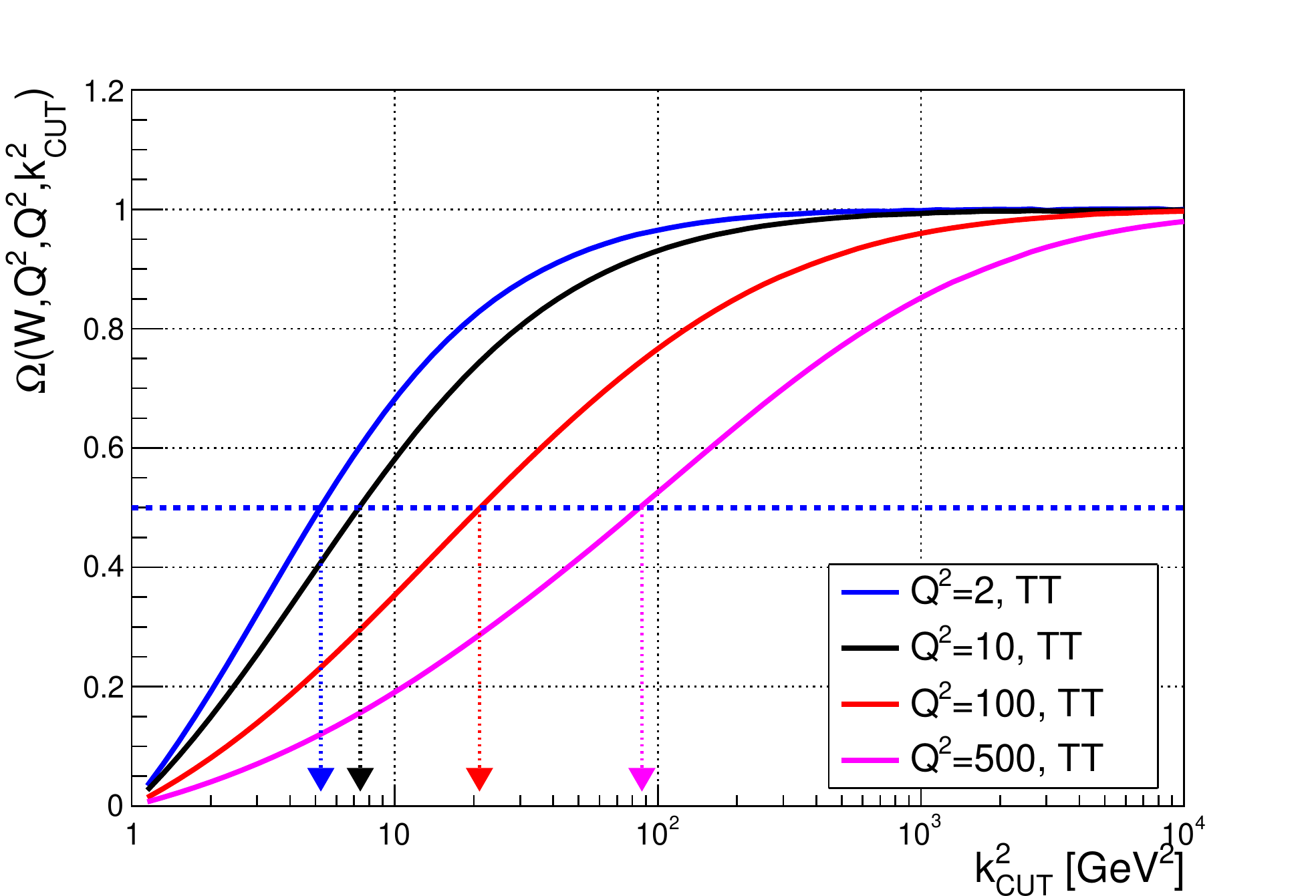}
	\includegraphics[width=0.49\textwidth]{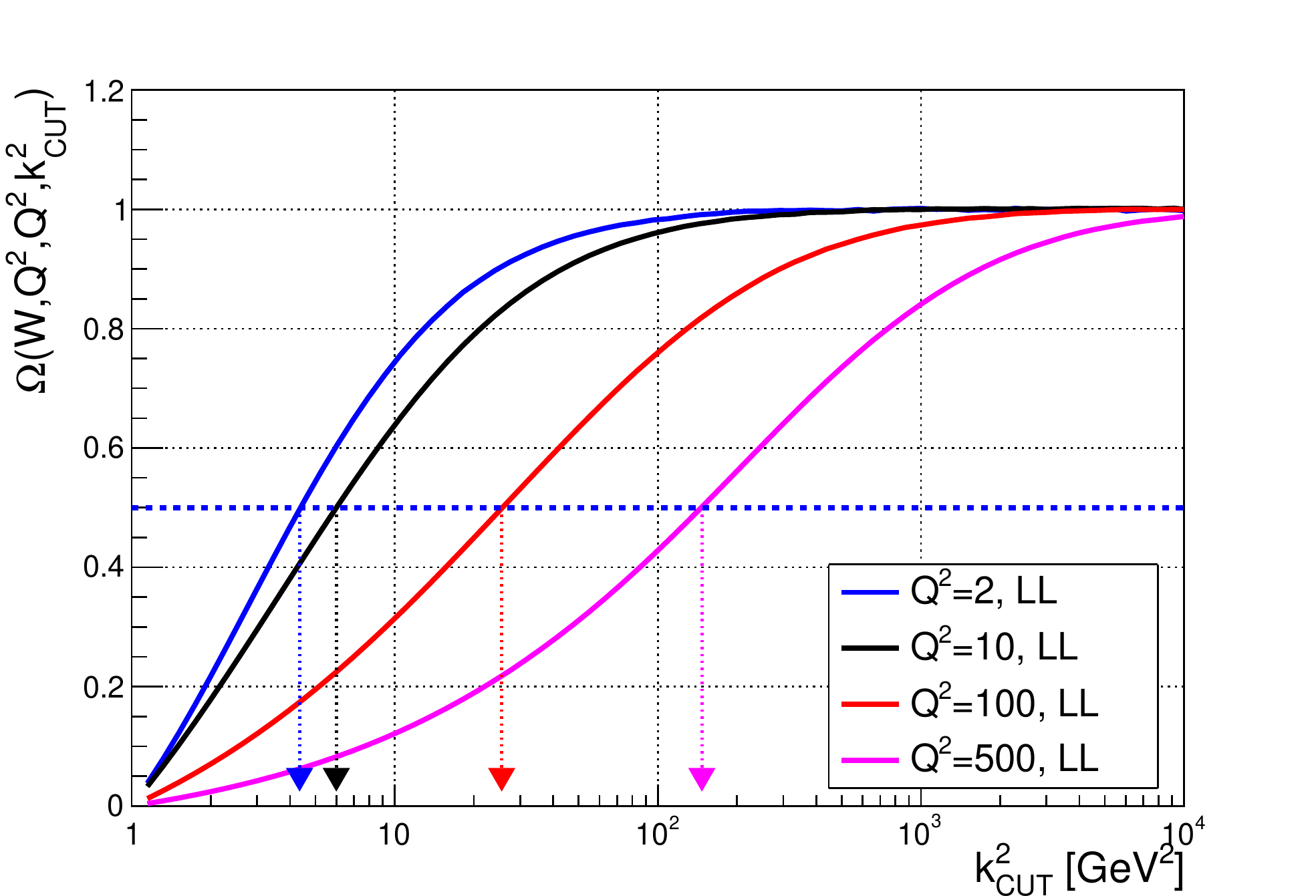}
	\caption{The ratio of the cross section for $\gamma^* \gamma^*$ scattering integrated up to a cut off $k_{\scaleto{\rm CUT}{4pt}}$ over the cross section integrated over the entire range of transverse momenta as a function of $k_{\scaleto{\rm CUT}{4pt}}$. The arrow indicate the positions of the values of $k_{\scaleto{\rm CUT}{4pt}}$  --median values -- for which the ratio is equal to $1/2$. Different curves indicate the different values  of $Q^2$: $2,10,100,500 \; {\rm GeV}^2$ -- blue, black, red and magenta lines correspondingly. The calculation was done for running  coupling constant, both in impact factors and in the BFKL evolution   and for the case of longitudinal-longitudinal (right plot) and transverse-transverse (left plot) polarization of both photons. Only light quark contribution is included in the calculation.}
	\label{fig:ktmedian_run}
\end{figure}

\begin{figure}
	\includegraphics[width=0.49\textwidth]{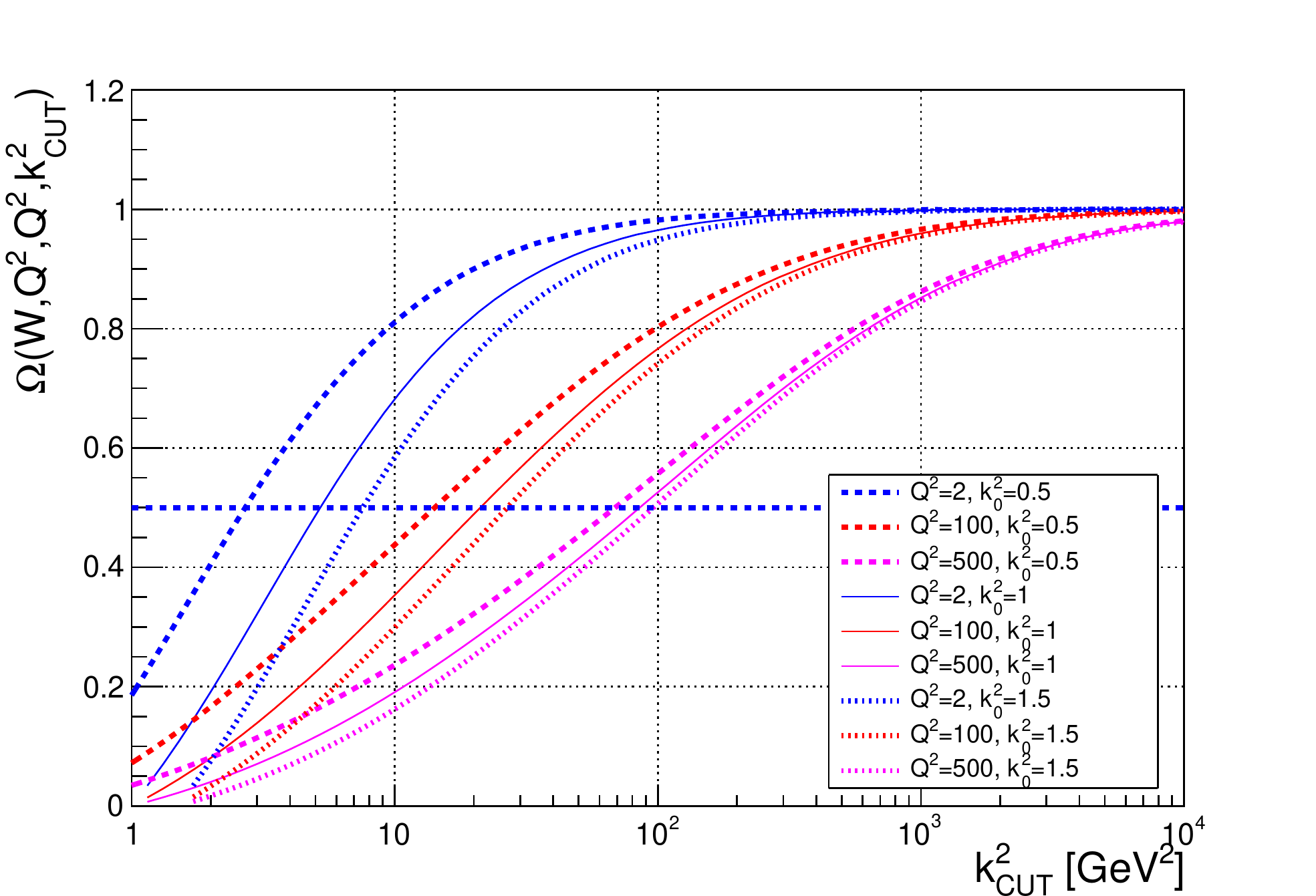}
	\includegraphics[width=0.49\textwidth]{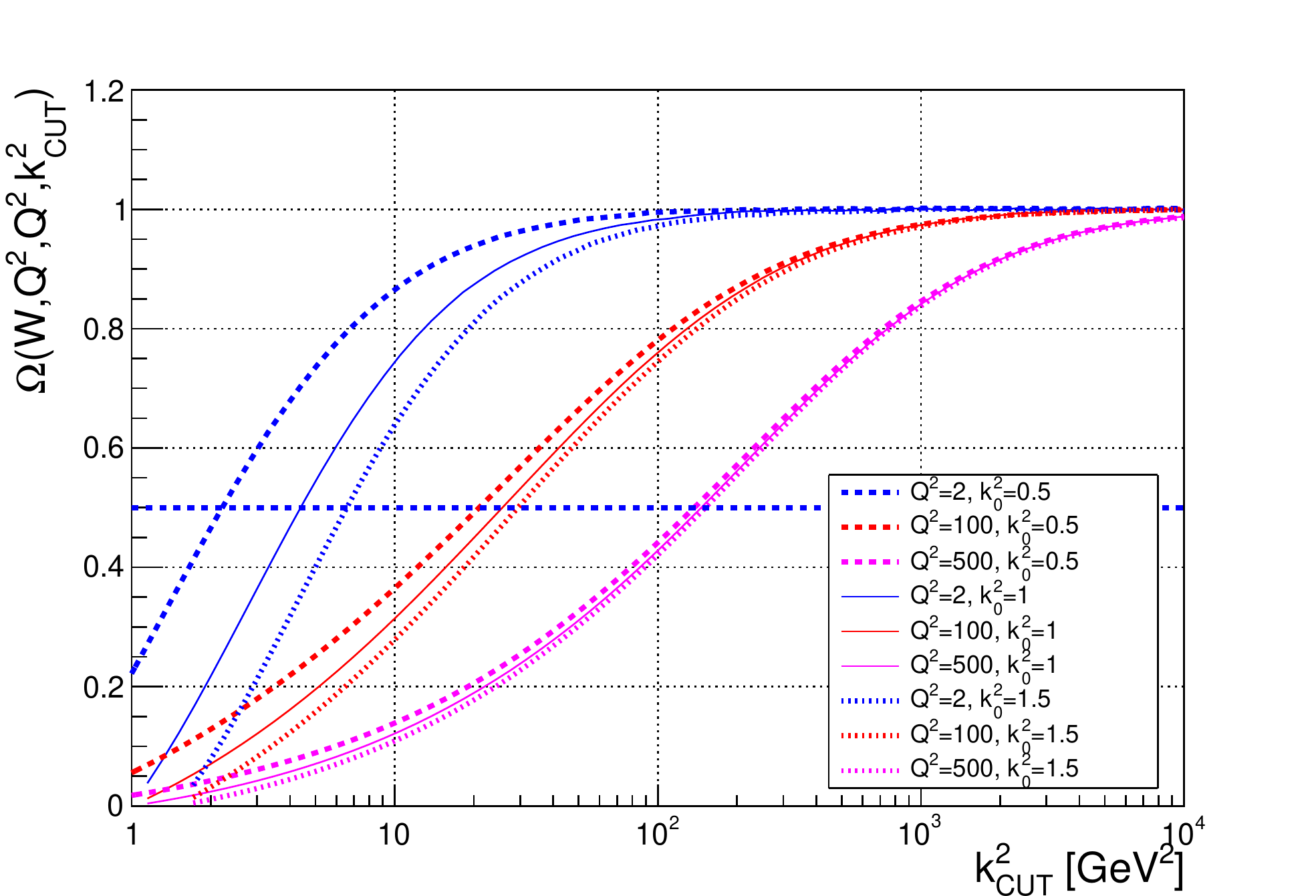}
	\caption{The ratio of the cross section for $\gamma^* \gamma^*$ scattering integrated up to a cut off $k_{\scaleto{\rm CUT}{4pt}}$ over the cross section integrated over the entire range of transverse momenta as a function of $k_{\scaleto{\rm CUT}{4pt}}$.  Different colored sets of curves indicate the different values  of $Q^2$: $2,100,500 \; {\rm GeV}^2$ -- blue, red and magenta lines correspondingly. The calculation was done for running  coupling constant, both in impact factors and in the BFKL evolution   and for the case of longitudinal-longitudinal (right plot) and transverse-transverse (left plot) polarization of both photons. Within the sets of lines, different curves indicate different non-perturbative cutoffs for the transverse momenta  $k_0^2=0.5, 1, 1.5 \; \rm GeV^2$.}
	\label{fig:ktmedian_run_cutoff}
\end{figure}
The relative importance of the contributions from the different $k^2$ regions is also illustrated in Figs.~\ref{fig:ktmedian_fxd} and \ref{fig:ktmedian_run} where we show the following quantity
\begin{equation}
{ \Omega}(W,Q_1^2,Q_2^2,k_{\scaleto{\rm CUT}{4pt}}^2) \; = \;  \int_{k_0^2}^{k_{\scaleto{\rm CUT}{4pt}}^2} \frac{dk^2}{k^2} { p}_{ij}(Q_1^2,Q_2^2,W,k^2) \;,
\label{eq:kmedian}
\end{equation}
as a function of the cutoff $k_{\scaleto{\rm CUT}{4pt}}^2$. This ratio clearly illustrates the  typical transverse momenta which are contributing to the cross section. 
The arrows indicate the median value of the $k_{\scaleto{\rm CUT}{4pt}}$. The Figs.~\ref{fig:pintegrand_fxd}, \ref{fig:pintegrand_run} can be interpreted as the logarithmic derivatives
over $k_{\scaleto{\rm CUT}{4pt}}^2$ of results  shown in Figs.~\ref{fig:ktmedian_fxd} and \ref{fig:ktmedian_run}.
In the fixed coupling case the median is very closely located to the value of the $Q^2$, for moderate to high values of $Q^2=10,100,500 \, \rm GeV^2$. In the case of the running coupling, for $Q^2=10,100,500 \, \rm GeV^2$, there is significant shift of the median to much lower values of $k^2$, which is more pronounced for the transverse-transverse case.  The exception is the case when $Q^2$ is very small, i.e. $Q^2=2 \, \rm GeV^2$. In this case the median seems to be shifted to  higher values of $k^2$, even in the case of the running coupling. This is of course a very low scale, close to the non-perturbative cutoff,
and the dependence on the lower cutoff $k_0^2$ should be significant. To test the sensitivity of the calculations to this parameter, we have performed the calculations with $k_0^2=0.5 \, \rm GeV^2$ and $k_0^2 = 1.5 \, \rm GeV^2$ in addition to $k_0^2=1 \, \rm GeV^2$. We show the results in Fig.~\ref{fig:ktmedian_run_cutoff}.  Again, left is transverse-transverse   and right plot is for the longitudinal-longitudinal case. For clarity, we only show three values of $Q^2=2,100,500 \, \rm GeV^2$.
We see significant cutoff dependence, as expected for low $Q^2=2 \, \rm GeV^2$. The cutoff dependence becomes  smaller for $Q^2=100 \, \rm GeV^2$ and is almost negligible for higher values. It is also clear that longitudinal-longitudinal case is somewhat less dependent on the value of the non-perturbative cutoff than the transverse-transverse calculation, which is expected  effect, due to the smaller width of the $k_T$ distribution.

\section{Summary and Conclusions}

Within QCD we found that the small $x$ resummation approach which accounts for the leading and non-leading logarithmic terms to small $x$ phenomena leads to a new portrait of perturbative  Pomeron.
The contribution of preasymptotic terms for one hard scale processes is relevant for the diminishing of effective intercept in a wide kinematical range and also  for the suppressing of onset of BFKL Pomeron
at moderate $x$, or equivalently at moderate energies. We showed that the onset of the BFKL growth can be delayed by $4-8$ units of rapidity for transverse scales between $3-100 \; {\rm GeV}$. The analysis indicates approximately logarithmic dependence of the `dip' (or the width of the plateau) in the gluon Green's function on the transverse momentum $k$.
This, rather complicated, energy dependence of the amplitude indicates that in a wide   preasymptotic region of $x$ the resummed  Pomeron  for 
one hard scale processes looks  as the sum of contributions  with different energy behaviors. 

We also found that the asymptotic value of the intercept is achieved only for very large rapidities, of the order of $20$ units  of rapidity or larger. On top of that, we found that even for the smallest values of the coupling constant, where the resummation effects should be the smallest and one would naively expect the dominance of the leading logarithmic approximation, the preasymptotic effects are still  important for very large $y$ values and in practice the solution is never close to the leading logarithmic approximation even at the highest energies and very small values of the coupling constant. This most likely stems from the fact that the LL approximation violates energy momentum conservation which cannot be remedied by the perturbative expansion in the strong coupling constant.

As an example of the application of the resummed calculation to the physical process   we calculated cross section of the process $\gamma^{*}+\gamma^*\to X$ which is very well suited for the hunt for perturbative Pomeron. 
We found that the cross section at lowest $\gamma^*\gamma^*$ energies is suppressed with respect to the Born calculation which is just the two gluon exchange. The suppression extends to energies of the order of $W \sim 100 \; \rm GeV$. Since we observe no such effect in the case of the calculation based on the LLx BFKL evolution we conclude that this is most likely related to the strong preasymptotic features found in the analysis of the resummed gluon Green's function. The BFKL growth can still  be observed, and leads to the enhancement  over the Born cross section by a factor $2-3$ for energies above $200 \; \rm GeV$. This onset of the asymptotic behavior strongly depends on the values of $Q^2$, and is delayed to higher energies for higher values of $Q^2$.  Similarly to the analysis of the gluon Green's function the cross section with the resummed BFKL solution is much lower than the calculation with LLx solution, factor of about 5 for energies of $1000 \; \rm GeV$, for high scales $Q^2 \simeq 100 \; \rm GeV^2$.

We have also studied the contributions to the cross section stemming  from the different regions of the transverse momenta of the gluon density. The integrand in the transverse momentum exhibits the maximum, however its position is shifted towards the infrared with respect to the external scale $Q^2$. This is simply related to the running of the coupling and to the fact that the distributions in the transverse momenta are relatively broad, especially for the transverse-transverse configurations, due to the nature of the impact factors. The distribution in the transverse momentum is additionally broadened by the BFKL diffusion. For example, for scales $Q^2=500$ and $100 \; \rm GeV^2$, the median transverse momentum is shifted to about  $80$ and $15 \; \rm GeV^2$ for the transverse-transverse case and to $150$ and $20 \; \rm GeV^2$ for the longitudinal-longitudinal configuration.

We expect that the phenomena that we have discussed here are quite general and hence would manifest themselves in other processes which are described in terms   of  single as well as multiple vacuum exchange.

\section*{Acknowledgments}
We thank Leszek Motyka for discussions.
This work  was supported by the  Department of Energy Grants No. DE-SC-0002145 and DE-FG02-93ER40771, as well as the National Science centre, Poland, Grant No.\ 2015/17/B/ST2/01838.


\bibliography{mybib}
\bibliographystyle{unsrt}

\end{document}